\documentstyle[12pt,aasms4]{article}

\def\be{\begin{equation}}
\def\ee{\end{equation}}
\def\ba{\begin{eqnarray}}
\def\ea{\end{eqnarray}}

\def\<{\noindent }

\newenvironment{namelist}[1]{%
\begin{list}{}
  {
   
   \settowidth{\labelwidth}{#1}
   \setlength{\leftmargin}{1.1\labelwidth}
  }
 }{%
\end{list}}

\def\o{\omega}
\def\f{\phi}
\def\n{\nu}
\def\p{\psi}
\def\a{\alpha}
\def\t{\theta}
\def\r{\frac{1}{r}}
\def\sr{\frac{1}{r^2}}
\def\nn{\nonumber}
\def\oh{\frac{1}{2}}

\journalid{VOL}{JOURNAL DATE}
\articleid{START PAGE}{END PAGE}
\paperid{MANUSCRIPT ID}

\slugcomment{Submitted to the Astrophysical Journal}

\lefthead{Stergioulas, Friedman}
\righthead{Nonaxisymmetric Neutral Modes}

\begin{document}

\title{Nonaxisymmetric Neutral Modes of Rotating Relativistic Stars}
\author{Nikolaos Stergioulas, John L. Friedman}
\affil{University of Wisconsin-Milwaukee, P.O. Box 413, Milwaukee, WI
53201 
 \\ niksterg@eexi.gr, friedman@thales.phys.uwm.edu}

\begin{abstract} 
We study nonaxisymmetric perturbations of rotating relativistic stars.
modeled as perfect-fluid equilibria.  Instability to a mode with
angular dependence $\exp(im\phi)$ sets in when the frequency of the
mode vanishes.  The locations of these zero-frequency modes along
sequences of rotating stars are computed in the framework of general
relativity.  We consider models of uniformly rotating stars with
polytropic equations of state, finding that the relativistic models are
unstable to nonaxisymmetric modes at significantly smaller values of
rotation than in the Newtonian limit.  Most strikingly, the m=2 bar
mode can become unstable even for soft polytropes of index $N \leq
1.3$, while in Newtonian theory it becomes unstable only for stiff
polytropes of index $N \leq 0.808$.  If rapidly rotating neutron stars
are formed by the accretion-induced collapse of white dwarfs,
instability associated with these nonaxisymmetric, gravitational-wave
driven modes may set an upper limit on neutron-star rotation.
Consideration is restricted to perturbations that correspond to polar
perturbations of a spherical star.  A study of axial perturbations is
in progress.

\end{abstract}


\section{Introduction}
\label{s:introduction}

Rotating perfect-fluid equilibria are unstable to nonaxisymmetric
instabilities driven by gravitational radiation.  Although apparently
damped by viscosity in neutron stars cold enough to have a superfluid
interior, the instability may play a role in limiting the maximum
rotation of newly formed neutron stars.  In particular, if rapidly
rotating neutron stars with weak magnetic fields form in the
accretion-induced collapse of some white dwarfs (or the core-collapse
of some stars), the instability may set an upper limit on rotation more
stringent  than the Kepler frequency -- the frequency of a satellite in
circular orbit at the star's equator. 
 
Prior to the present work, the onset of the nonaxisymmetric
instability had  been computed only in the Newtonian limit and
estimated in the first post-Newtonian approximation and in a
slow-rotation approximation.\footnote {After this was written, we
received a preprint by Yoshida and Eriguchi (1997) computing
instability points using a Cowling approximation in general
relativity.  This is an approximation that ignores changes in the
gravitational potential, and it should be accurate for large-m modes.}  The
first fully relativistic computation is presented here (and in Stergioulas'
PhD thesis (1996)).  

That stars can be unstable to gravitational radiation was first found
by \markcite{Chandra70} Chandrasekhar (1970), who considered the $m=2$
modes for Maclaurin spheroids (uniform density rotating stars) in a
Newtonian context.  \markcite{F&S75,F&S78a,F&S78b} Friedman \&
Schutz (1975, 1978a, 1978b), show that this instability also
appears in compressible stars and that all rotating ``stars'' (rotating
self-gravitating perfect-fluid configurations) are generically unstable
to the emission of gravitational radiation.  Even for slowly rotating
models there will always be a polar mode of high enough mode number $m$
(equivalently, of short enough wavelength) that is unstable.  
For a perfect fluid, a nonaxisymmetric mode becomes unstable when its
frequency vanishes in the inertial frame, i.e., with respect to an
observer at infinity. This occurs when a star rotates fast enough that
an inertial observer sees that a counterrotating mode becomes
corotating with the star. The amplitude of the perturbation then grows
with time, making the star unstable.
 
Realistic neutron stars are viscous, and the presence of viscosity will
shift the onset of instability. Work by Lindblom \& Mendell
(1995)\markcite{LM95} (see also Ipser \& Lindblom 1991a,b;  Yoshida \&
Eriguchi 1995)\markcite{IpserLindblom91a,IpserLindblom91b,YoshidaEriguchi95} shows that viscosity will damp out the instability once
the temperature is low enough that the interior is superfluid.  A
window apparently remains, of temperatures less than about $10^9$ K and
greater than about $10^{10}$ K, where the instability will affect a
sufficiently rapidly rotating neutron star as it cools down after its
birth.  In the Newtonian limit, it appears that only the $l=m \leq 5$
modes are not damped out by viscosity and the $l=4$ mode apparently
sets the most stringent limit on the maximum angular velocity; the
critical angular velocity is about 90 \% to 95 \% of the Kepler limit.
Old neutron stars, spun-up by accretion, may be too cold to be subject
to the gravitational radiation driven instability.

A surprise, recently pointed out by Andersson (Andersson 1997; Friedman
and Morsink 1997)\markcite{Andersson 1997,F&M97} is that axial modes
for all values of $m$ will be unstable for perfect-fluid models with
arbitrarily slow rotation.  In a spherical star, axial perturbations
are time-independent convective currents that do not change the density
and pressure of the star and do not couple to gravitational waves.  For
rotating stars, their growth time $\tau$ is proportional to a high
power of $\Omega$ (rough arguments appear to imply
$\tau\propto\Omega^{-4-2m}$), and viscosity will again presumably
enforce stability except for hot, rapidly rotating neutron stars.

The onset of the nonaxisymmetric instability for several modes in
rotating polytropes has been computed in the Newtonian limit by
Imamura, Friedman \& Durisen (1985) \markcite{IFD85} using a
Langrangian variational principle and by Managan (1985)
\markcite{Managan85} and Ipser \& Lindblom (1990)
\markcite{IpserLindblom90} using an Eulerian variational principle
constructed by Ipser \& Managan (1985) \markcite{IpserManagan85}.
Cutler (1991) \markcite{Cutler91}, Cutler \& Lindblom (1992)
\markcite{CutlerLindblom92} and Lindblom (1995) \markcite{Lindblom95}
estimate the correction to first post-Newtonian order. Weber et al
(1991) \markcite{WGW91} provide an estimate in a relativistic,
slow-rotation approximation.  Since neutron stars are relativistic
objects, relativity must have a significant effect on the onset of
instability. The post-Newtonian analysis suggests that, for a given
equation of state (EOS) and a given mode $l$, the ratio
$\Omega_c/\Omega_k$ of the critical angular velocity where the mode
becomes unstable to the maximum allowed angular velocity (the Kepler
frequency $\Omega_k$) decreases as the neutron star becomes more
relativistic.  Thus, nonaxisymmetric instabilities set a more stringent
limit on the maximum angular velocity than Newtonian theory suggests.

In this paper, we report the first computation of zero-frequency modes
of rotating relativistic stars.  Polytropes of index $N=1.0$, 1.5 and
2.0 are considered, and attention is restricted  to the fundamental (f)
$l=m \leq 5$ modes, since these are the ones that are most likely to
participate in limiting the maximum angular velocity of neutron stars.
It is shown that the critical angular velocities of relativistic stars
are considerably lower than the corresponding Newtonian estimates. As a
result, the $m=2$ bar mode can become unstable for much softer
($N<1.3$) polytropes than in the Newtonian limit.  This is expected to
hold true for most realistic EOSs since they have an effective index
$N<1.0$. Thus, depending on the equation of state and the effect of
viscosity, the $m=2$ mode may participate in setting the upper limit on
rotation for rapidly rotating neutron stars created by the
accretion-induced collapse of white dwarfs.

In forthcoming work we plan to include the effect of viscosity on the 
critical angular velocities and extend our method to $N<1.0$ polytropes
and to realistic equations of state.


\section{The Equilibrium Configurations}
\label{s:equilibrium}

The spacetime of a rotating relativistic star in equilibrium can be
described by the stationary, axisymmetric metric of the form

\be ds^2 = -e^{2 \nu} dt^2 + e^{2 \psi} \bigl(d \phi -\omega dt \bigr)^2
+
             e^{2 \alpha} \bigl(dr^2+r^2 d \theta^2 \bigr),
\label{metric}
\ee
  
\<(for a review see Friedman \& Ipser (1992) \markcite{FriedmanIpser92}).

In (\ref{metric}), as in the rest of this paper, gravitational units 
($c=1$, $G=1$) are employed. The metric involves four independent
equilibrium metric potentials $\nu, \psi, \alpha$ and $\omega$ which
are  functions of $r$ and $\theta$ only. We assume uniform rotation,
 $\Omega=$ constant, where $\Omega$ is the angular velocity of the
star. The matter is assumed to be a perfect fluid at zero temperature,
described by a polytropic equation of state, for which the energy
density $\epsilon$, pressure $P$ and number density $n$ satisfy the
relations

\ba
         P   &=& K n^{1+1/N}, \label{pressure} \\ 
    \epsilon &=& n m_{\rm B} +NKn^{1+1/N}, \label{epsilon} 
\ea

\<where $K$ is a constant, $m_{\rm B}$ is the rest mass per baryon and
$N$ is the polytropic index, related to the adiabatic index $\gamma$ by 
\be 
    \gamma= \frac{d \ln P}{d \ln n} = \frac{\epsilon + P}{P} 
              \frac{dP}{d\epsilon} = 1+ \frac{1}{N}.  \label{Gamma}
\ee

The stress-energy tensor for a perfect fluid is

\be  T^{ab} = (\epsilon + P) u^a u^b + P g^{ab} \label{Tab}, \ee

\<where the equilibrium 4-velocity $u^a$ is given by

\be u^a = \frac{e^{-\nu}}{\sqrt{1-v^2}} \bigl( t^a + \Omega \phi^a 
          \bigr), \label{ua} \ee

\<with 

\be v= \bigl( \Omega - \omega \bigr) e^{\psi - \nu}, \label{v} \ee

\<the fluid velocity  with respect to a local zero-angular-momentum
observer. We have denoted by $t^a$ and $\phi^a$ the Killing vectors
$\partial_t$ and $\partial_phi$ associated with the time and rotational symmetries of the metric.

An equilibrium model satisfies the field equation, 
\be 
R_{ab} = 8 \pi \Bigl( T_{ab}- \frac{1}{2}g_{ab}T \Bigr), 
   \label{Fieldeqn} 
\ee
and the implied equation of energy conservation 
\ba 
  u_b \nabla_a T^{ab} = 0 \Longrightarrow 
\ \ u^a \nabla_a \epsilon = -(\epsilon+ P)  \nabla_a u^a.
  \label{energy}
\ea
and Euler equation
\be
  q^c{}_b \nabla_a T^{ab} = 0 \Longrightarrow 
  \ \ u^a \nabla_a u^b = \frac{q^{bc} \nabla_c P}{\epsilon + P},
  \label{Euler}
\ee
Here $q^{ab}=g^{ab}+u^a u^b$ is the projection of the
metric
tensor orthogonal to the 4-velocity. 

\<Numerical equilibrium models are constructed by a 
numerical code (Stergioulas \& Friedman (1995)
\markcite{StergioulasFriedman}) that implements the Cook, Shapiro \&
Teukolsky (1994) \markcite{CST94} version of the KEH method (Komatsu,
Eriguchi \& Hachisu (1989) \markcite{KEH89}). Our code has been shown
to be accurate in an extensive comparison with other existing codes
(Eriguchi {\it et al.} (1996) \markcite{comparison} ). The field
equations are solved on a 2-dimensional grid that is uniform in $\mu
\equiv \cos \theta$ and in the radial coordinate

\be s=r/(r+r_e), \label{sdefinition} \ee
 
\<which compactifies the region $r=0$ to $+\infty$ to the finite
region $s=0$ to $1$ ($r_e$ is the value of the coordinate $r$ at the
equator).
This is important, because the gravitational potentials have a
nonnegligible value even far from a relativistic star; and it also
simplifies the implementation of boundary conditions in the
construction of both the equilibrium and the perturbed configurations.
Note that the equator of the star is always at $s=0.5$, so that (for a
nonrotating star) half of the radial grid-points are inside the star
and the other half are outside.

We define dimensionless quantities by setting $G=c=1$  and by fixing the 
length scale. If we define

\be \tilde{K} = \frac{K}{m_{\rm B}^{1+1/N}}, \ee
 
\<then $\tilde{K}^{N/2}$ can be used as the fundamental length scale. 
Dimensionless quantities will be denoted as $\bar{\epsilon}$,
$\bar{\Omega}$
etc. The dimensionless energy density and angular velocity are then

\be \bar{\epsilon} = \frac{\tilde{K}^N G}{c^4} \epsilon, \ee

\<and

\be \bar{\Omega}= \frac{\tilde K^{N/2}}{c} \Omega. \ee

\<In subsequent sections, numerical results will be reported using 
dimensionless quantities as defined above. {\em Note that the dimensionless
quantities are independent of the polytropic constant $K$.}


\section{The Perturbed Configurations}
\label{s:perturbed}

We will use an Eulerian formalism \markcite{FriedmanIpser92} to describe fluid
perturbations.  The Eulerian perturbation in the metric tensor,
 
\be \delta g_{ab} \equiv h_{ab}, \label{hab}, \ee
 
\<can be determined by solving the perturbed field equations,
 
\be \delta R_{ab} = 8 \pi \bigl( \delta T_{ab} - \frac{1}{2} g_{ab}
\delta T -
                      \frac{1}{2} h_{ab} T \bigr), \label{dField} \ee
 
\<in a suitably chosen gauge. The perturbation in any other quantity is
evaluated to first order in $h_{ab}$. The change in the Ricci tensor
is then (see e.g. Wald 1984 \markcite{Wald84})

\be \delta R_{ab} = \nabla_c \nabla_{(a} h^c{}_{b)} - \frac{1}{2} \Bigl(
    \nabla_a \nabla_b h_c{}^c + \nabla^c \nabla_c h_{ab} \Bigr), \ee

\<where $h^a{}_b = g^{ac} h_{cb}$. Priou 1992 \markcite{Priou92} has 
explicitly computed the components of $\delta R_{ab}$ for a stationary, 
axisymmetric background in its most general form
(prior to any choice of gauge). This allows us to directly use Priou's
results
with only the following modifications: 
\\ i) since we are only interested in zero-frequency modes, we set the
frequency of the mode and all time-derivatives equal to zero;\\
ii) we rename Priou's functions, changing $m$ to $M$, $l$
to $L$ and $\Omega$ to $y$, in order to avoid confusion with the indices $l,m$ that characterize a mode and
with the angular velocity $\Omega$ of the equilibrium star; and \\
iii) we choose a gauge as described in section \ref{s:gauge}.\\ 
The components of $\delta R_{ab}$ in our gauge are displayed in Appendix 
\ref{a:Ricci}.

The perturbation of the stress-energy tensor is

\be \delta T_{ab} = u_a u_b ( \delta \epsilon + \delta P) + ( \epsilon
+ P)
(u_a \delta u_b + u_b \delta u_a) + g_{ab} \delta P + P h_{ab}, 
 \label{dTab}, \ee

\<while the perturbation of its contraction is

\be \delta T = 3 \delta P - \delta \epsilon, \label{dT} \ee
  
\<The change in pressure and energy density can be expressed in
terms of a single scalar function $\delta U$, defined implicitly by the
relation

\be
\delta P = (\epsilon + P)  \bigl ( \delta U + {1 \over 2} u^a u^b 
             h_{ab} \bigr ). \label{dP} 
\ee
For an adiabatic perturbation, we have 
\be
\delta \epsilon = {(\epsilon + P) \over P \Gamma} \delta P,
         \label{de} 
\ee

\<and we will assume that the 
adiabatic index $\Gamma$ of the perturbation be equal to the 
adiabatic index $\Gamma_0$ of 
the equilibrium configuration, defined in (\ref{Gamma}).

In the perturbed relativistic Euler equations,
\be
\delta (q^a{}_c \nabla_b T^{bc}) = 0,
\label{dq}
\ee
the only derivatives of $\delta u^a$ that occur are along the unperturbed
fluid velocity $u^a$.  For perturbations with harmonic $t$- and 
$\phi$-dependence, the equations are thus algebraic in $\delta u^a$ and
can be solved analytically (Ipser \& Lindblom, 1992)
\markcite{IpserLindblom92}.  The perturbed fluid velocity is given by
 
\be
\delta u^a = i \ Q^{ab} \Bigl[ \nabla_b \bigl( \delta U + {1 \over 2} u^c
u^d 
              h_{cd} \bigr ) -  \delta F_b \Bigr]+ {1 \over 2} u^a u^c
u^d 
              h_{cd}, \label{dua}
\ee
     
\<where

\be \delta F_a =  q_a{}^b \Biggl[ - u^c \nabla_c ( h_{bd} u^d) +u^c u^d 
\nabla_b 
                  h_{cd}+u^c u^d h_{cd} {\nabla_b P \over \epsilon + P} 
\Biggr]. 
     \label{dFa} \ee
   
\<The tensor $Q^{ab}$ involves only equilibrium quantities and the
frequency $\sigma$ of the mode in a rotating frame; for the 
time-independent perturbations considered here,

\be \sigma = m\Omega . \label{sm} \ee
Explicitly,
\be Q^{ab} = {1 \over D} \left[ (\sigma u^t)^2q^{ab}-2\omega^a\Omega^b 
+ i\sigma u^t
(2\hat\phi^b\epsilon^{ac}\Omega_c-\hat\phi^a\epsilon^{bc}\omega_c)\right]
. \label{qab} \ee 
Here
\be u^t = u^a\nabla_at = \left[ e^{2v}-e^{2\psi}(\Omega -\omega
)^2\right]^{1/2} ; \label{utua} \ee 
\<the quantity
\be D=(\sigma u^t)^3 -2\sigma u^t\Omega^a\omega_a \label{dsu} \ee
is the determinant of $Q^{-1}$; $\omega^a$ is the vorticity of the fluid,

\be \omega^a = \epsilon^{abcd}u_b \nabla_c u_d; \ee

\<$\Omega^a$ is a generalization of the angular-velocity vector, 

\be \Omega^a = \frac{1}{2} u^t \epsilon^{abcd} u_b (\nabla_c t_d +
\Omega 
    \nabla_c \phi_d); \ee

\<$\hat\phi^a$ is the  unit linear
combination
of the Killing fields that is orthogonal to $u^a$,

\be \hat{\phi}^a = \frac{e^{-\psi -v}}{u^t} q^a{}_b\phi^b; \ee

\<and the
tensor
$\epsilon^{ab}$ is the volume element on the 2-surfaces 
orthogonal to 
the Killing trajectories

\be \epsilon^{ab} = \epsilon^{abcd} \hat{\phi}_c u_d. \ee 
 
The
tensor 
$Q^{ab}$ exists (and the perturbed Euler equations can be inverted) as
long
as the determinant in (\ref{dsu}) does not vanish.
 
      For completeness, we list components of $u_a$, $\omega^a$ and
$Q^{ab}$.  With $u^t$ given by Eq. (\ref{utua}), we have
\begin{eqnarray*}
u_\phi &=& u_a\phi^a = e^{2\psi}(\Omega -\omega )u^t ;\\
\omega^r &=& \frac{e^{-\psi -\nu -2\alpha}}{r(u^t)^2} \partial_\theta
(u^tu_\phi )\\
\omega^\theta &=& - \frac{e^{-\psi -\nu -2\alpha}}{r(u^t)^2} \partial_r
(u^t u_\phi )\\
Q^{tt} &=& \sigma^2e^{-2\psi -2\nu} D^{-1}(u^tu_\phi )^2\\
Q^{tr} &=& -Q^{rt} =-i\sigma e^{-\psi -\nu}r\omega^\theta
D^{-1}u^t u_\phi\\
Q^{t\theta} &=& -Q^{\theta t} =-i\sigma e^{-\psi -\nu} r^{-1}\omega^r
D^{-1}u^tu_\phi\\
Q^{t\phi} &=& Q^{\phi t} = \frac{1+\Omega u^tu_\phi}{u^tu_\phi} Q^{tt}\\
Q^{rr} &=& D^{-1} [(\sigma u^t)^2e^{-2\alpha}-(\omega^r)^2]\\
Q^{r\theta} &=& -Q^{\theta r} = -D^{-1}\omega^r\omega^\theta\\
Q^{r\phi} &=& -Q^{\phi r} = i\sigma e^{-\psi -\nu} r\omega^\theta
D^{-1}(1+\Omega u^tu_\phi)\\
Q^{\theta\theta} &=& r^{-2}D^{-1}[(\sigma u^t)^2 e^{-2\alpha} -
r^2(\omega^\theta )^2]\\
Q^{\theta\phi} &=& -Q^{\phi\theta} =-i\sigma e^{-\psi -\nu} r^{-
1}\omega^r D^{-1}(1+\Omega u^tu_\phi)\\
Q^{\phi\phi} &=& \left( \frac{1+\Omega u^tu_\phi}{u^tu_\phi}\right)^2
Q^{tt}
\end{eqnarray*}

\<The perturbed covariant fluid velocity is given by

\be \delta u_a=\delta u^b g_{ab} +u^b h_{ab}. \label {du_a} \ee

Through (\ref{dP}), (\ref{de}), (\ref{dua}), (\ref{dFa}) and the
equation (\ref{du_a}) for the perturbed covariant fluid velocity, the
perturbation in the stress-energy tensor $\delta T_{ab}$ is expressed
entirely in terms of the perturbed metric $h_{ab}$ and the scalar
$\delta U$.  Due to the lengthy substitutions involved, we used the
algebraic program MAPLE to compute $\delta u_a$ which was then
substituted in $\delta T_{ab}$.  The expressions for the components of
$\delta T_{ab}$ in the stationary, axisymmetric background are several
pages long (for each component), and we do not display them here. These
expressions, along with the components of $\delta R_{ab}$, are used to
form the perturbed field equations (\ref{dField}).  MAPLE is used to
convert these expressions into numerical code for inclusion in an
ANSI-C program.

The perturbed field equations determine $h_{ab}$ for given $\delta U$.
The scalar function $\delta U$ is determined by an additional equation,
the perturbed energy conservation equation

\be \delta(u_b \nabla_a T^{ab}) =0. \label{d_en_cons} \ee

\<which will be considered in section \ref{s:energy_eqn}.


\section{Expansion in Spherical Harmonics}
\label{s:expansion}

Because the equilibrium configuration is axisymmetric, linear
perturbations of the star and geometry can be decomposed into a sum of
terms with angular dependence $e^{im\phi}$.  We expect that, as is the
case for spherical stars, each discrete mode of a nonrotating, spherical
model has a continuous extension to a mode for each rotating model with
the same equation of state.  Modes of spherical stars have angular
dependence given by the tensor, vector and scalar harmonics associated
with a given $Y_l^m(\theta ,\phi )$.  The corresponding mode of a
rotating model can thus be labeled by $l$ and $m$; but all harmonics
$Y^m_{l'}$, with $l' \geq l\geq m$ contribute to the mode.

Studies of rotating Newtonian stars find nonaxisymmetric instability sets
in first along a sequence with increasing rotation for a 
reflection-invariant polar mode with $l=m$, the lowest value of $l$ for
a given $m$.  We have, in this first numerical study, correspondingly
restricted our consideration to perturbations invariant under reflection
in the equatorial plane (under the diffeo $\theta \rightarrow\pi -
\theta$){\footnote {Recall that if a vector $v^a$ is invariant under the
diffeo $\theta\rightarrow\pi -\theta$ its components $v^r=v^a\nabla_ar$,
$v^t=v^a\nabla_at$, and $v^\phi = v^a\nabla_a\phi$ are invariant, while
$v^\theta =v^a\nabla_a\theta$ changes sign because $\nabla_a\theta$
changes sign.}}.

The Eulerian perturbation in a reflection-symmetric, $l=m$, 
time-independent mode has the form

\ba
h_{tt}, h_{rr} & \propto & \sum_{l'=0}^{\infty} a_{m+2l'} \ Y_{m+2l'}^m
\\
h_{tr}& \propto & \sum_{l'=0}^{\infty} i \ a_{m+2l'+2} \ Y_{m+2l'+2}^m
\\
h_{t \mu} & \propto & \sum_{l'=0}^{\infty} \Bigl( \ i \ a_{m+2l'+2} \ 
               \Psi_{m+2l'+2
            \ \mu}^m + b_{m+2l'+1} \ \Phi_{m+2l'+1 \ \mu}^m \Bigr) \\
h_{r \mu} & \propto & \sum_{l'=0}^{\infty} \Bigl( \ a_{m+2l'+2} \ 
           \Psi_{m+2l'+2      
             \ \mu}^m + i \ b_{m+2l'+1} \ \Phi_{m+2l'+1 \ \mu}^m \Bigr)
\\
h_{\mu \nu} & \propto & \sum_{l'=0}^{\infty} \Bigl(\ a_{m+2l'} \
\Phi_{m+2l' \ 
                \mu \nu}^m + b_{m+2l'+2} \ \Psi_{m+2l'+2 \ \mu \nu}^m  
           \nonumber \\ 
            & & \ \ \ \ \ \ \  + \  i \ c_{m+2l'+1} \ \chi_{m+2l'+1 \ \mu
\nu}^m 
            \Bigr)
\ea

\<where the $a$'s, $b$'s, and $c$'s are real coefficients, different for
each component of $h_{ab}$.  We have adopted the Regge-Wheeler (1957)
notation for spherical harmonics:

\begin{eqnarray}
\Psi_l^m{}_\theta\/ &= \partial_\theta Y_l^m, \qquad\qquad\qquad\qquad\qquad \Psi_l^m{}_\phi &=
\partial_\phi Y_l^m ;\\
\Phi_l^m{}_\theta &= - \frac{1}{\sin\theta} \partial_\phi Y_l,
\qquad\qquad\qquad\qquad \Phi_l^m{}_\phi &=\sin\theta \partial_\theta Y_l^m;\\
\Phi_l^m{}_{\theta\theta} &= Y_l^m, \quad\qquad\qquad\qquad\qquad\qquad\Phi_l^m{}_{\theta\phi}&=0;\\
\Phi_{l\phi\phi}^m &= \sin^2\theta Y_l^m;\qquad\qquad\qquad\qquad\quad \qquad&\\
\Psi_l^m{}_{\theta\theta} &= \partial_\theta^2Y_l^m,\qquad\qquad\qquad\qquad\qquad 
\Psi_l^m{}_{\theta\phi} &= (\partial_\theta\partial_\phi -
\cot\theta\partial_\phi )Y_l^m,\\
\Psi_l^m{}_{\phi\phi} &= (\partial^2_\phi
+\sin\theta\cos\theta\partial_\theta )Y_l^m;\quad\qquad\qquad&\qquad\\
\chi_l^m{}_{\theta\theta} &= \frac{1}{\sin\theta} (\partial_\theta -
\cot\theta )\partial_\phi Y_l^m, \qquad\qquad \chi_l^m{}_{\theta\phi} &=
\frac{1}{2} \left( \frac{1}{\sin\theta} \partial^2_\phi
+\cos\theta\partial_\theta -\sin\theta\partial^2_\theta\right) Y_l^m,\\
\chi_l^m{}_{\phi\phi} &= -\sin\theta (\partial_\theta -\cot\theta
)\partial_\phi Y_l^m .\qquad\qquad\quad&
\end{eqnarray}

In Priou's (1992) \markcite{Priou92} notation (and with our own 
redefinitions mentioned 
in section \ref{s:perturbed}), the perturbed metric $h_{ab}$ is expressed 
in terms of ten $\phi$-dependent {\it perturbation functions}, $h,p,k,w,q,a,b,L,M,$ and $y$, in the manner
  
\ba
h_{tt} & = & -2he^{2 \nu} + (2 y \omega + 2w \omega^2)e^{2 \psi}   \\
h_{tr} & = & L+a \omega e^{2 \psi}     \\
h_{t \theta} & = & M+b\omega e^{2 \psi}     \\
h_{t \phi} & = & -e^{2 \psi}(y +2 \omega w)     \\
h_{rr} & = & 2ke^{2 \a}     \\
h_{r \theta} & = & q     \\
h_{r \phi} & = & -ae^{2 \psi}     \\
h_{\theta \theta} & = & 2pr^2e^{2 \a}     \\
h_{\theta \phi} & = & -be^{2 \psi}     \\
h_{\phi \phi} & = & 2we^{2 \psi},     
\ea

\<Equivalently, to first order in $h_{ab}$

\begin{eqnarray} 
 ds^2 &=& -e^{2 \nu}(1+2h)dt^2 
     + e^{2 \psi}(1+2w) \Bigl[ d \phi - (\omega+y) dt 
   - \ a dr -b d \theta \Bigr]^2 \nonumber \\
 & & +e^{2 \a} \Bigl[ (1+2k)dr^2  
     + r^2(1+2p) d \theta^2 \Bigr] \nonumber \\ 
    & &  + \ 2qdrd \theta + 2Ldtdr + 2Mdtd \theta.
\end{eqnarray}
 
An associated set of real, $\phi$-independent, variables $\hat L,\cdots
\hat w$, can be defined by writing

\ba
\hat h &=& he^{-im\phi} , \\
\hat k &=& ke^{-im\phi} , \\
\hat L &=& - i \ L \ e^{-im\phi} , \label{97} \\
\hat M &=& - i \ M \sin \theta \ e^{-im\phi} ,\\
\hat q &=& q \sin \theta \ e^{-im\phi} , \\
\hat y &=& y \sin^2 \theta \ e^{-im\phi} ,\\
\hat a &=& -i \ a \sin^2 \theta \ e^{-im\phi} ,\\
\hat b &=& -i \ b \sin^3 \theta \ e^{-im\phi} ,\\
\hat p &=& p \sin^2 \theta \ e^{-im\phi} , \label{predef} \\ 
\hat w &=& w \sin^2 \theta \ e^{-im\phi} . \label{wredef}
\ea
 
\<Then, the metric perturbations of these modes can be expanded as follows
in terms 
of Legendre polynomials:
 
\begin{eqnarray}
\hat h,\  \hat k,\ \hat L,\ \hat y,\ \hat a,\ \hat p,\ \hat w   &\sim&  \sin^m
\theta \ 
\sum_{l'=0}^{\infty} a_{2l'}(r) P_{2l'}(\cos \theta), \label{evenrefl}
\\
 \hat M,\ \hat q,\ \hat b   &\sim&   \sin^m \theta \ \sum_{l'=0}^{\infty}
a_{2l'+1}(r) P_{2l'+1}(\cos \theta) \label{oddrefl}, 
\end{eqnarray} 
 
\<It is obvious from (\ref{evenrefl}) and
(\ref{oddrefl}), that the functions $\hat M, \hat q$ and $\hat b$ will
be
antisymmetric about the equatorial plane, while all other functions
will be symmetric. Furthermore, with these definitions, 
{\it the time-independent perturbed field equations become a system of 
ten real equations for ten real unknowns}.

The corresponding expansions for axial $l=m$ modes, for which the
tensor $h_{ab}$ changes sign under reflection (modes that reduce to
axial modes in the Regge-Wheeler gauge in the spherical limit), can
easily be written in a similar fashion. We note that the
behavior under reflection in the equatorial plane will be opposite 
to that of polar modes.

\section{Gauge Choice}
\label{s:gauge}

A linear
perturbation of a star, described by the set of quantities $(\delta
\epsilon$,
 $\delta p$, $\delta u^a$, $h_{ab}$) is physically equivalent to the
gauge-related perturbation described by the set 
$(\delta \epsilon + {\cal L}_{\eta} \epsilon$, $\delta p+ {\cal L}_{\eta}
p$, 
$\delta u^a + {\cal L}_{\eta} u^a$, $h_{ab}+ {\cal L}_{\eta} g_{ab} $)
for any smooth vector field $\eta^a$ that preserves the asymptotic behavior
of the metric.  The choice of gauge is important
for the successful numerical solution of the perturbed field equations.
After experimenting with a
large 
number of
possible gauges, we found that a numerical solution was more easily
obtained
in a gauge defined by the four conditions

\ba
   h_{r \theta} &=& 0    \ \ \  \Longrightarrow  q  =0, \\
   h_{\theta \phi} &=& 0 \ \ \   \Longrightarrow b  =0, \\
   h_{t \phi} &=& - \omega h_{\phi \phi}  \ \ \ \Longrightarrow  y=0, \\
   h_{\phi \phi} &=& \frac{h_{\theta \theta}}{r^2}e^{2(\psi-\a)} \ \ \
   \Longrightarrow  w=p.
   \label{hphiphi}
\ea

The components of $h_{ab}$ satisfy all imposed boundary conditions in
this gauge.  In addition, we required that it reduce to the
Regge-Wheeler polar gauge in the non-rotating limit, in which only
$h_{tt}$, $h_{rr}$, $h_{\theta \theta}$ and $h_{\phi \phi}$ are
non-zero, with $h_{\theta \theta}$ and $h_{\phi \phi}$ satisfying the
equivalent of (\ref{hphiphi}) in a Schwarzschild metric (see  previous
section). With this choice of gauge there are six nonzero metric
functions in the list (\ref{97})--(\ref{wredef}), namely $h$,
$p$, $k$, $\hat L$, $\hat M$ and $\hat a$.  They will be
expressed in terms of the function $\delta U$ by solving six components
of the field equation $\delta R_{ab} = 8\pi\delta (T_{ab}-\frac{1}{2}
g_{ab}T)$, $(tt)$, $(rr)$, $(\theta\theta )$ $(tr)$, $(t\theta )$ and
$(r\theta )$.  Note that condition (\ref{hphiphi}) implies that the
perturbation function $p$ does not have to be redefined, as in
(\ref{predef}), because it has the desired angular behavior

\be
 p \ \ \sim \ \ \sin^m \theta \ \sum_{l'=0}^{\infty} a_{2l'}(r) P_{2l'}
 (\cos \theta).
\ee

In Appendix \ref{a:Ricci} we list the necessary components of the
perturbation 
in the Ricci tensor in this gauge.


\section{Numerical Solution}
\label{s:numerical}
The task of solving the coupled system of six differential equations is
not
trivial.
In the gauge specified in section \ref{s:gauge} the $(tt)$ and 
$(\theta \theta)$ equations are elliptic for h and p respectively. The 
$(rr)$ equation is parabolic for k (it is missing a 
$\partial^2k/\partial r^2$ derivative). The $(tr)$ and $(r \phi)$
equations
are second order ordinary differential equations (ODEs) for $L$ and $a$ 
in the angular direction, while the $(t \theta)$ equation is a 
second order ODE for $M$ in the radial direction. Each type of equation
requires its own finite-difference scheme and boundary conditions.
Still, we were able to solve all six equations simultaneously on a 
2-dimensional 
finite grid.

In the radial direction, the grid coincides with that of the
equilibrium configuration; that is, it consists of a number of
grid-points, uniformly spaced in the coordinate $s$, which ranges from
$s=0$ $(r= 0)$ to $s=1$ $(r=\infty)$, the equator always being at
$s=0.5$ (the coordinate $s$ is defined by (\ref{sdefinition})).  In the
angular direction a different grid is used.  Following Ipser \&
Lindblom (1990) \markcite{IpserLindblom90}, the $n$ angular grid-points
are located at the angles $\mu_i =\cos \theta_i$ which correspond to
the zeros of the Legendre polynomial of order $2n-1$:
$P_{2n-1}(\mu_i)=0$. {\em This has the advantage of using only a small
number of angular grid-points to describe the star, which results in a 
smaller linear system of equations to be solved.}
Since the perturbation functions
are either symmetric or antisymmetric when reflected in the equatorial
plane, one need only solve for $0< \theta \leq \pi/2$, and the boundary
conditions at $\theta=\pi/2$ are incorporated in the expressions for
the angular derivatives.  A drawback of this method is that stars with
stiff equations of state $(N<1.0)$ are less accurately described than
stars with soft EOSs $(N>1.0)$, because the former have discontinuous
derivatives of energy density and metric functions across the surface
and this results in the appearance of Gibbs phenomena in the angular
derivatives.
  
The above choice of angular grid-points allows us to use high-order
formulae
for the angular derivatives. In all six equations, all first and second
order angular derivatives are approximated as

\be \frac{\partial}{\partial \mu} f^{\pm}(s_i, \mu_j) = 
   \sum^n_{k=1} D_{jk}^{\pm} f^{\pm}(s_i, \mu_k), \label{dm} \ee

\<and

\be \frac{\partial^2}{\partial \mu^2} f^{\pm}(s_i, \mu_j) = 
   \sum^n_{k=1} H_{jk}^{\pm} f^{\pm}(s_i, \mu_k), \label{dmm} \ee

\<where $f^{\pm}$ is a function which is either symmetric $(+)$ or 
antisymmetric $(-)$ under the transformation $\theta \rightarrow \pi
-\theta$, 
i.e. $f^{\pm}(s,-\mu)= \pm f(s,\mu)$. The functions $D_{jk}^{\pm}$ and
 $H_{jk}^{\pm}$
are derived in Appendix \ref{a:derivative} and are constructed
specifically 
for 
functions that have the angular behavior

\be f^+= \sin^m \theta \sum_{l'=0}^{\infty} a_{2l'} P_{2l'}(\mu), \ee
 
\<like $h,k,p,\hat L$ and $\hat a$, and
 
\be f^-= \sin^m \theta \sum_{l'=0}^{\infty} a_{2l'+1} P_{2l'+1}(\mu), \ee

\<like $\hat M$.

In all equations, except for the parabolic equation $(rr)$, the first and
second order radial derivatives are approximated by the standard, second
order accurate, central difference formulae

\be \frac{\partial}{\partial s} f_{i,j}= \frac{f_{i+1,j}-f_{i-1,j}}{2
\Delta 
s}, 
   \label{ds} \ee

\<and

\be \frac{\partial^2}{\partial s^2} f_{i,j}= \frac{f_{i+1,j}-2f_{i,j}+
f_{i-1,j}}
    {\Delta s^2}, 
 \label{dss} \ee

\<where $\Delta s$ is the distance between two radial grid-points.
In the parabolic equation $(rr)$, we use an implicit solution scheme;
that is,
we use a first order accurate, backwards, first radial derivative for k,
   
\be \frac{\partial}{\partial s} f_{i,j} = \frac{f_{i,j}-f_{i-1,j}}{\Delta
s}, 
\ee

\<while for the first order derivatives of other functions involved in
the
$(rr)$ equation we still use the second order accurate equation 
(\ref{ds}).
Second order radial derivatives in the $(rr)$ equation are not
approximated
by (\ref{dss}) but by a similar equation using twice the grid-spacing
 
\be \frac{\partial^2}{\partial s^2} f_{i,j}= \frac{f_{i+2,j}-2f_{i,j}+
f_{i-2,j}}
       {(2 \Delta s)^2}.
 \label{dss2} \ee

\<If one tries to use (\ref{dss}), the solution for $k$ oscillates. That 
using (\ref{dss2}) instead of (\ref{dss}) can suppress such oscillations
was discovered in the construction of equilibrium models, where a second
order 
radial derivative appears in the source term of a non-elliptic equation,
by Stergioulas \& Friedman (1995) \markcite{StergioulasFriedman}.
Mixed derivatives involved in some of the equations are approximated by

\be
\frac{\partial^2}{\partial s \partial \mu}f_{i,j}^{\pm} = \frac{1}{\Delta
s} 
\sum_{k=1}^n D_{jk}^{\pm} \bigl( f_{i,k}-f_{i-1,k}). \ee

\<At the center and at infinity the 
appropriate boundary finite difference formulae are used in all
equations. 

For the two elliptic equations $(rr)$ and 
$(\theta \theta)$ we require that the solution has vanishing first order
angular derivative at $\theta=\pi/2$ and vanishes at infinity.
The boundary condition at $\theta=\pi/2$ is built into the
construction of the angular derivative formulae. No boundary conditions
are needed at the center and on the symmetry axis for the elliptic
equations,
but the discretized equations  force the solution to
vanish 
there for any mode with $m \neq 0$, as is the case for the exact
solution.

For the two second order angular ODE's, $(tr)$ and $(r \phi)$, the
boundary
conditions at $\theta=\pi/2$ and on the symmetry axis are set by the
angular
derivative equations (\ref{dm}) and (\ref{dmm}). For the second order
radial 
ODE $(t \theta)$ we require that $\hat M$ vanishes at the center and at
infinity. Finally, for the parabolic equation $(rr)$, boundary conditions
at $\theta=\pi/2$ and on the symmetry axis are set by the angular
derivative
equations (\ref{dm}) and (\ref{dmm}), while in the radial direction we
set
$k=0$ at the center only (at infinity the boundary is open).

Of the above equations, the most difficult to solve numerically are the 
$(rr)$ and $(r \phi)$ equations. In the $(rr)$ equation, the solution is 
propagated from the center to infinity via a first order radial
derivative,
so it is far more sensitive to local inaccuracies (like those 
produced by the finite grid-spacing at the surface) than the elliptic
equations. In particular, the surface of the rotating star does not
correspond to a constant value of the radial coordinate but falls in
between
grid points. For polytropes of index $N \geq 2$, the energy density goes
smoothly enough to zero at the surface that this does not pose any
problem
even with a grid of only 200 radial points. For polytropes of index
$1 \leq N <2$, however, a small jump in the solution for $k$ appears at
the
surface. Stiff polytropes of index $N<1$ have, in addition,  
discontinuous  second order derivatives of the equilibrium metric
functions 
across the surface and a jump in the numerical solution for $k$ is
unavoidable
when using our grid, since the angular derivative equations were designed
only
for smooth functions. The small jump appearing in the numerical solution
for
$k$ does not affect any other perturbation function significantly, except
for 
the function $a$. Another problem with the $(rr)$ equation - of same
origin - 
is that, for stiff equations of state, the solution oscillates in the
vacuum 
region, unless one makes the approximation

\be \frac{\partial^2 h}{\partial s^2} +2 \frac{\partial^2 p}{\partial
s^2}
    \simeq -\frac{\partial^2 h}{\partial s^2}. \ee
 
\<The error introduced by making this approximation is small, since
for all stars that we examined
 
\be h \simeq -p. \label{hpksim}\ee
 
\<Thus, the significant benefit of making this approximation 
(suppression of all oscillations in the solution) outweighs
its mild cost. 
 
Finally, the $(r \phi)$ equation seems to be very sensitive to local
inaccuracies, especially at the surface of the star and near its
center.  The inaccuracies near the center of the star occur, because
the perturbation function $a$ depends on the {\it differences} between
other perturbation functions, which all have very small values
(compared to their maximum value) near the center.  However, the
perturbation function $a$ is of lesser importance compared to the other
five functions, when computing the critical angular velocities. In
fact, we determined that $a$ only minimally affects the other five
perturbation functions, so that setting it equal to zero, results in
almost unchanged solutions for the other perturbation functions and
unchanged critical angular velocities for even the most relativistic
stars of the stiffest polytropic index $(N=1.0)$ we examined.

We also determined that the other two off-diagonal perturbation functions
$\hat L $ and $\hat M$ are weakly coupled to the diagonal ones. This
is expected, since in our gauge the perturbed metric reduces to a
diagonal
form in the non-rotating limit and  $h_{t \phi}$ is the
dominant off-diagonal perturbation ($g_{t \phi}$ is the only
non-vanishing off-diagonal metric element for a rotating star). We 
emphasize 
that all the above statements hold for perturbations {\em of the type
considered in the present paper}.
 
The finite-differencing of the system of perturbed field equations
yields a large linear system. The unknowns are ordered in a way that
casts the matrix of the linear system in a band-diagonal form. A direct
solution method (LU-decomposition) is employed. Alternatively, we also
use an iterative Bi-Conjugate Gradient method with a {\em symmetric succesive
over-relaxation (SSOR) method} as a preconditioner. The direct solution was 
faster, but required a larger amount of random access memory.\footnote{The
subroutines for the solution of the linear system we used, are part of
the Portable Extensible Toolkit for Scientific Computation (PETSc)
package developed at Argonne National Laboratory (Gropp and Smith 1994)
\markcite{GroppSmith94}.}
  
A representative solution of the coupled system of six perturbation
equations,
for a specific trial function $\delta U$, is shown in Figs. (\ref{fig1})
- 
(\ref{fig3}).
The background star is a rapidly rotating, N=1.5 polytrope with
dimensionless 
central energy density $\bar{\epsilon}_c=0.061$ and dimensionless angular
velocity $\bar{\Omega}=0.092$. The solution was obtained for the $l=m=3$
mode
using a trial function

\be \delta U = r^l P_l^m (\mu), \ee
 
\<where $P_l^m(\mu)$ is the associated Legendre function. This trial
function is the dominant term in  the true eigenfunction of the $l=m$
modes (see section \ref{s:energy_eqn}). All six equations were solved
on the same $ 101 \times 7$ (radial $\times$ angular) grid. The
perturbation functions $h$, $p$, $k$, $\hat L$, $\hat M$  and $\hat a$
are shown along all seven angular spokes and at the pole (where they
vanish). With the exception of $\hat M$, the perturbation functions are
symmetric under reflection in the equatorial plane and their maximum
value occurs for $\theta= \pi/2$. The perturbation function $\hat M$ is
antisymmetric under reflection, so it vanishes at $\theta=\pi/2$. Its
maximum value occurs about half-way between the pole and the equator.
The diagonal perturbation functions $h$, $p$ and $k$ have no nodes
inside the star. This is a characteristic of $h_{tt}$ for fundamental
$l=m$ modes in the Newtonian limit. The perturbation function $k$ (the
solution to the parabolic $(rr)$ equation) exhibits a small jump at the
surface, as was explained earlier. In addition, near the surface, the
solution can have a wave-like character. This is normal and arises
because the solution is obtained for a trial function $\delta U$ and
{\it not} for the true eigenfunction. 
The off-diagonal functions $\hat L$, $\hat M$ and $\hat a$
vanish much faster in the vacuum surrounding the star than the diagonal
functions do; they become negligible almost immediately outside the
surface of the star. The function $\hat L$ does not have any node
inside the star, but $\hat M$ and $\hat a$ do. 
Finally, the solution for the function $\hat a$ exhibits
some oscillations at the surface of the star, which are induced by the
behavior of the function $k$ there.  Regarding other polytropic indices
and modes, the perturbation functions remain similar to those in Figs.
(\ref{fig1}) - (\ref{fig3}). For stiffer polytropes and for more
relativistic stars, the solutions are peaked closer to the surface than
for softer polytropes and less relativistic stars. For larger $m$, the
solutions are more narrowly peaked about their maximum value because
the dominant radial behavior of $\delta U$ is $r^l$. For $N=1.0$
polytropes the oscillations at the surface are larger than for $N=1.5$
polytropes, while for $N=2.0$ polytropes they are negligible.
Otherwise, the solutions for the perturbation functions are very
similar to those in Figs. (\ref{fig1}) - (\ref{fig3}).

\section{A Truncated Gauge}
\label{s:truncated}
 
In the Newtonian limit, the important component of the perturbation
$h_{ab}$
is $h_{tt}$, which reduces to $-2 h$. In addition, the 
metric functions $h$, $k$, $p$ and $w$ satisfy the relation

\be h=-k=-p=-w. \label{hkpw} \ee

\<For energy densities typical in neutron stars, the $h_{tt}$ component
still
dominates $h_{ab}$. We determined numerically, that even for the most 
relativistic
and rapidly rotating $N=1.0$ polytrope, the relation (\ref{hkpw}) still
holds
approximately in our gauge. This led us to investigate whether by making 
certain 
approximations, one can still obtain accurate values for the critical 
angular velocities of neutral modes, while solving fewer than six
equations. 
Indeed,
we find that by making the approximations

\be \frac{h_{tt}}{g_{tt}+2 \omega g_{t \phi}} = \frac{h_{rr}}{g_{rr}} 
     = \frac{h_{\theta \theta}}{g_{\theta \theta}}  
    \ \ \ \   \Longrightarrow \ \ \ \  h=-k=-p, \label{approx1} \ee
  
\<and

\be h_{t \theta} = h_{r \phi}  = 0 \  \ \ \  
    \Longrightarrow \ \ \ \ \hat{M} = \hat{a} =0, 
     \label{approx2} \ee

\<one can obtain very accurate critical angular velocities, while
solving only two perturbed field equations. Equation (\ref{approx1}) is
motivated by (\ref{hkpw}) and (\ref{approx2}) is used since $g_{t
\theta}=g_{t \phi}=0$ for the equilibrium configuration and because
$h_{t \theta}$, $h_{r \phi}$ vanish in the spherical limit in the
Regge-Wheeler polar gauge , i.e. they are proportional to the
background metric function $\omega$.  In this {\it truncated gauge},
the perturbation in the metric tensor, can be written in the form
 
\be h_{ij}=
\pmatrix{-2h(e^{2 \nu}-\omega^2 e^{2 \psi}) & L & 0 & -2 \omega h 
            e^{2 \psi} \cr   
              & -2 h e^{2 \alpha}  & 0 &  0         \cr
          {\rm sym.}     &                &  -2h r^2e^{2 \alpha} &  0  
   \cr 
              &     &      &  -2h e^{2 \psi}   \cr}. \ee 
 
\<Only two perturbation functions are nonzero, $h$ and $\hat L$, which
are 
determined by solving the $(tt)$ and $(tr)$ equations.

Table \ref{t:trunc} shows a comparison of critical configurations
obtained in the full and truncated gauges. The equilibrium star is a
$N=1.0$ polytrope with $\bar{\epsilon}_c =0.3$, which is close to the
central energy density of the maximum mass configuration. The critical
configurations for the $l=m=3,$ 4 and 5 modes are computed in both
gauges on the same $ 101 \times 7$ grid.  We compare the dimensionless
critical angular velocity and the critical ratio of rotational to
gravitational binding energy $T/|W|_c$ in the two gauges. The critical
values of $T/|W|$ differ by less than $1 \%$ and the critical angular
velocities by less than $0.5 \%$ when computed in the truncated gauge
compared to computing them in the full gauge.  The advantage of using
the truncated gauge is that far less memory is needed by the numerical
code. While six equations can be solved in a reasonable time with a
maximum $201 \times 12$ grid on a DEC Alpha with 256 MBytes of memory,
two equations can be solved with a much finer grid of $ 801 \times 12$
or more points. The execution speed in not limited by the processor
speed (the matrix inversion is very fast) but by the available random
access memory.  For very soft polytropes of index $N \geq 2.0$, a $ 201
\times 12$ grid gives sufficiently accurate critical angular
velocities, so that one can use the full gauge for these models. But,
for polytropes of $N \leq1.0$ a $201 \times 12$ grid determines the
critical angular velocity with an error that can exceed several percent
or more. This was determined by comparing critical configurations for
Newtonian polytropes, obtained with our code, to published results.
Thus, for realistic neutron stars, it is necessary to obtain the
critical angular velocities with a grid of at least $801 \times 12$
points and this could be achieved only in the truncated gauge. For this
reason, 
{\em all results reported in the present paper} were obtained in the
truncated gauge with a grid of $801 \times 12$ points, unless
otherwise stated.


\section{The Perturbed Energy Conservation Equation}
\label{s:energy_eqn}

In preceding sections it was shown how a solution of the perturbed
field equations is obtained for a given trial function $\delta U$.  The
complete description of the neutral modes requires the satisfaction of
both the perturbed field equations (\ref{dField}) and the perturbed
energy conservation equation (\ref{d_en_cons}) by the perturbation in
the metric $h_{ab}$ and the scalar function $\delta U$.  This system of
equations (\ref{dField}) and (\ref{d_en_cons}) has a zero-frequency
solution only for the stars for which a zero-frequency mode exists. For
any equilibrium model, however, one can solve only the perturbed field
equations (\ref{dField}) for trial functions $\delta U$ and use the
perturbed energy conservation equation (\ref{d_en_cons}) to construct a
criterion for locating the marginally stable star (for a given mode) along
a sequence of rotating stars.

The perturbed field equations (\ref{dField}) are implicitly linear in the
function $\delta U$, as is the perturbed energy conservation equation
(\ref{d_en_cons}). Hence (\ref{d_en_cons}) can be
represented as a linear operator $L$ acting on a function $\delta U$

\be L(\delta U) = 0. \label{LdU}\ee
 
\<Equation (\ref{LdU}) represents an eigenvalue problem for the linear 
operator $L$ with zero eigenvalue and eigenfunction $\delta U$.

In the Newtonian limit, it was shown that the eigenfunction $\delta U$
of an $l=m$ neutral mode can be approximated accurately by expanding it
in terms of a  set of basis functions $\delta U_i$,

\be \delta U = \sum_i  a_i \delta U_i, \label{dU_exp} \ee
 
\<of the form 
\markcite{IpserLindblom90,SkinnerLindblom96}
 
\be \delta U_i = \delta U^{(jk)}_i = r^{l+2(j+k)} Y_{l+2k}^m (\cos
\theta), 
\label{dUi} \ee
  
\<with $j,k=0,1,...$ and with each set of indices $(j,k)$ yielding 
a particular $\delta U_i$. In practice 
the eigenfunction $\delta U$  is represented with
reasonable
accuracy by only a few terms, and the $(0,0)$ term $r^l Y_l^m$ dominates the
expansion
(\ref{dUi}).

\<Substituting the expansion (\ref{dU_exp}) in (\ref{LdU}) yields

\be \sum_i a_i L(\delta U_i) =0. \label{SLdU} \ee

If we define the inner product\footnote{The motivation for defining the
inner 
product as in (\protect{\ref{innerproduct}}) will become apparent in 
Appendix \protect{\ref{a:energy}}.} 

\be < \delta U_j | L | \delta U_i > \ \ = \ \ \int i \frac{\delta U_j}{\sigma
u^t} 
 L(\delta U_i) \sqrt{-g} d^3x,
     \label{innerproduct} \ee

\<where $g$ is the determinant of the equilibrium metric tensor, then
taking 
the inner product of (\ref{SLdU}) with respect to $\delta U_j$ yields

\be \sum_i a_i < \delta U_j | L | \delta U_i > =0. \label{SdULdU} \ee

\<Although the basis functions are not orthogonal to each other, non-zero
$< \delta U_j | \delta U_i >$ terms do not appear on the r.h.s. of 
(\ref{SdULdU}) because we require that 
the eigensystem has zero eigenvalue.
The last equation is a linear, homogeneous system for the coefficients
$a_i$.
The system has a non-trivial solution only if its determinant vanishes.
This
yields a criterion for locating the zero-frequency modes

\begin{namelist}{Criterionxx}
\item[{\bf Criterion}] {\it A stationary, axisymmetric model has a 
nonaxisymmetric, zero-frequency mode with angular dependence $e^{im \phi}$, 
if

\be {\rm det} < \delta U_j | L | \delta U_i > = 0. \label{det} \ee }

\end{namelist}

\<In practice, we start with a slowly rotating star and compute the
matrix elements $< \delta U_j | L | \delta U_i > $. For slowly rotating
stars the determinant in (\ref{det}) always has a large value. Keeping
the central energy density constant, we look at stars of increasing
angular velocity, until the determinant goes through zero. The star for
which (\ref{det}) is satisfied is the one for which the particular
$l=m$ mode has a zero-frequency solution and the nonaxisymmetric
instability sets in through that mode. In this method, the accuracy in
locating the neutral modes depends on how well the expansion
(\ref{dU_exp}) approximates the true eigenfunction $\delta U$.

In Appendix \ref{a:energy}, we obtain, in terms of Eulerian quantities,
an expression for $<\delta U_j | L | \delta U_i >$  that is used in the
numerical computations. As discussed in this Appendix the above method
of solving the perturbed energy conservation equation is not identical
to using a variational principle; because of our choice of field
equations, the matrix $< \delta U_j | L | \delta U_i >$ is not
symmetric.  For central energy densities typical in neutron stars,
however, the matrix is nearly symmetric and the method nearly coincides
with a variational principle.

\section{Critical configurations}
\label{s:results}

Following the method developed in previous sections, we computed the
neutral mode (critical) configurations for the fundamental $l=m=2,$ 3,
4 and 5 nonaxisymmetric modes. Three equations of state were examined,
having polytropic indices $N=1.0$, 1.5 and 2.0 polytropes. The $N=1.0$
polytropes include models with mass and radius similar to those of
realistic neutron stars.  The other two equations of state were
examined for completeness and for comparison with published results in
the Newtonian limit. The critical configurations were computed in the
truncated gauge with a fine grid of up to $801 \times 12$ (radial
$\times$ angular) grid points. The trial function $\delta U$ was
expanded as in (\ref{dUi}) using different number of terms. Eight basis
functions, corresponding to the indices $j=0,...,3$ and $k=0,1$ in
(\ref{dUi}) were sufficient to determine the critical configurations
with good accuracy, especially for the $N=1.5$ and $N=2.0$ polytropes,
where the error consistently decreases with increasing number of basis
functions and increasing number of grid points. This was not so for the
$N=1.0$ polytropes. Owing to the finite grid-size, the surface of the
star is discontinuous, and this gives rise to Gibbs phenomena at the
surface when using the expansions (\ref{dm}), (\ref{dmm}) for the
angular derivatives. As a result, increasing the number of grid-points
did not monotonically decrease the error. This behavior is expected for
our choice of coordinates and grid. It is the price one has to pay for
being able to solve a smaller than otherwise linear system of
equations.  A similar behavior is reported in Bonazzola {\it et al.}
(1996) \markcite{BFG96}, who use a similar angular grid in computing
the neutral, viscous bar mode in relativistic stars. Still, by
computing the $N=1.0$ configurations with various grid sizes and
various numbers of basis functions, we could obtain sufficiently
accurate results.

Our code was checked in the Newtonian limit by comparing the critical 
configurations for the $l=m=3$, 4 and 5 modes for $N=1.0,$ 1.5 and 2.0 
polytropes
to results published by Managan (1985) \markcite{Managan85}, Imamura, 
Friedman \& Durisen (1985) \markcite{IFD85} and Ipser \& Lindblom (1990)
\markcite{IpserLindblom90}.
As can be seen in Table \ref{t:Newtonian} we are in very good agreement
with 
all
published Newtonian results, the exception being $T/|W|_c$ for the $m=5$
mode
in $N=1.0$ polytropes, which differs by 2 \% from the value published in
Ipser \& Lindblom (1990) \markcite{IpserLindblom90}, from which Imamura, 
Friedman \& Durisen (1985) \markcite{IFD85} differ by 3 \% (in the
opposite 
direction). This highlights the
increased inaccuracy involved in computing neutral modes for $N \leq 1.0$
polytropes.

For relativistic polytropes we obtain the following results:  For $N=1.0$,
figure \ref{fig4} shows the ratio of the critical angular velocity
$\Omega_c$ to the {\it Keplerian angular velocity $\Omega_K$  at same
central energy density} as a function of central energy density for the
four modes examined. The lowest central energy density in the figure
corresponds to a mildly relativistic star.  The highest central energy
density shown is the central energy density of the most massive (and
thus most relativistic) star allowed by the particular equation of
state. The filled circles on the left vertical axes represent the
values of $\Omega_c/\Omega_K$ in the Newtonian limit.  As the central
energy density increases and the star becomes more relativistic,
$\Omega_c/\Omega_K$ decreases and it decreases at a faster rate as it
approaches the most relativistic configuration. Contrary to the
Newtonian limit, where $N=1.0$ polytropes do not have an unstable $m=2$
mode, we find that in relativistic $N=1.0$ polytropes the $m=2$ mode
becomes unstable when the central energy density exceeds roughly one
10th the central energy density of the most massive star. At the most
relativistic configuration, the $m=2$ mode becomes unstable for
$T/|W|_c$=0.065 or $\Omega_c/ \Omega_K=0.91$.  Table \ref{t:1.5}
compares the most relativistic critical configurations to their
counterparts in the Newtonian limit.  The value of $\Omega_c/\Omega_K$
decreases by roughly $15 \%$ for the $m=3$, 4 and 5  modes. Figure
(\ref{fig5}) shows the critical ratio $T/|W|_c$ for the same $N=1.0$
polytropes. This ratio decreases faster and by a larger percentage than
$\Omega_c / \Omega_K$, owing to the fact that the Keplerian value
$T/|W|_K$ also decreases as one samples more relativistic stars.  In
Table \ref{t:1.0} it is shown that for the most relativistic $N=1.0$
polytrope the critical ratio $T/|W|_c$ is about 40 \% smaller, for the
$m=3$, 4 and 5 modes, than the corresponding ratio in the Newtonian
limit. When the decrease in the Keplerian value $T/|W|_K$ is taken into
account and one looks at the ratio $(T/|W|_c) / (T/|W|_K)$ the most
relativistic values are still $25-30 \%$ lower than the Newtonian
values.

Figures (\ref{fig6}) and (\ref{fig7}) and Table \ref{t:1.5}
display 
the critical 
configurations for $N=1.5$ polytropes. These polytropes are softer than
the 
ones 
with index $N=1.0$. Consequently, they have a smaller maximum mass and
are less
relativistic. For this reason, relativity has a smaller effect on the
onset of
the nonaxisymmetric instability. The $m=2$ mode does not become unstable
even 
for
the most relativistic $N=1.5$ polytropes. For the $m=3$, 4 and 5 modes,
the 
value of $\Omega_c / \Omega_K$ 
decreases by $7-10 \%$ for the most relativistic models compared to the 
Newtonian limit.
The corresponding decrease for $T/|W|_c$ is $30-35 \%$ and for 
$(T/|W|_c) / (T/|W|_K)$ it is $13-19 \%$.
 
Plots of $\Omega_c / \Omega_K$ and $T/ |W|_c$ for the $N=2.0$ polytropes
are 
shown in figures (\ref{fig8}) and (\ref{fig9}). 
These are extremely soft models and their
maximum mass occurs at nearly Newtonian central energy densities. 
Again, the $m=2$ mode does not become unstable. As seen in table
\ref{t:2.0},
for the $m=3$, 4 and 5 modes the values of
$\Omega_c / \Omega_K$, $T/|W|_c$ and $(T/|W|_c) / (T/|W|_K)$ are only a
few
percent less than in the Newtonian limit, for the largest central energy 
density.
Polytropes constructed with $N=2.0$ do not resemble realistic neutron
stars,
but are included here for completeness. 

Regarding the accuracy of our results, we estimate that the determined 
critical angular velocities and critical $T/|W|_c$ are accurate to better than 
2 \% for $N=1.0$, 1-2 \% for $N=1.5$ and 1 \% for $N=2.0$ polytropes.

\section{Discussion}
\label{s:discussion}
 
We have treated nonaxisymmetric neutral modes in the context of general 
relativity.
We have found a gauge in which six coupled perturbed field equations can be 
solved
simultaneously with good accuracy. Furthermore, we found an
approximate gauge, in which the critical configurations for $N \geq 1.0$ 
polytropes are located with sufficient accuracy, while solving only two 
independent perturbed field equations. We showed that 
general relativity has a large effect on the location of nonaxisymmetric 
neutral modes and forces the nonaxisymmetric instability to set in for smaller
rotation rates than Newtonian theory suggests. 

The large effect of relativity on the onset of the nonaxisymmetric instability 
is most striking in the case of the $m=2$ modes. In the Newtonian context it 
was 
shown that uniformly or nearly uniformly  rotating neutron stars cannot become 
unstable to the $m=2$ mode unless the equation of state is excessively stiff
(see e.g. Skinner \& Lindblom (1996) \markcite{SkinnerLindblom96}). For 
polytropes, the classical result by James (1964) \markcite{James64} restricts 
the onset
of the m=2 instability to polytropes having an adiabatic index larger than
$\Gamma_{\rm crit} = 2.237$ (which corresponds to a polytropic index 
$N_{\rm crit}=0.808$). 
We find that, in the context of general relativity, this critical value 
becomes 
$\Gamma_{\rm crit} = 1.77$ (polytropic index $N_{\rm crit} = 1.3$). 
 
In the Newtonian limit, the $m=2$ mode driven by gravitational radiation
coincides with the m=2 mode driven by viscosity even for compressible fluids, 
as was shown by Ipser \& Managan (1985) \markcite{IpserManagan85}. 
It is interesting that work by Bonazzola {\it et al.} (1996) \markcite{BFG96}
suggests that in general relativity, the $m=2$ viscosity-driven mode has 
a critical adiabatic index only {\it slightly higher} than James's result.
Thus, in general relativity, the viscosity driven and the gravitational 
radiation
driven $m=2$ modes no longer coincide and the effect of relativity seems 
to be very different on each of them.

In forthcoming work, we plan to study realistic equations of state and include
the effect of viscosity on the onset of the nonaxisymmetric instability.

\acknowledgments

We would like to thank Lee Lindblom, Jim Ipser, Yoshiharu Eriguchi and 
Ed Seidel for very helpful discussions. This research has been supported by
NSF grants PHY-9105935 and PHY-9507740.


\appendix

\section{Perturbed Ricci Tensor}
\label{a:Ricci}
We list the perturbation in the six components of the Ricci tensor that are
used in the perturbed field equations, in the gauge of section \ref{s:gauge}.
We follow the notation introduced in section \ref{s:expansion}. In the
r.h.s. of each equation, subscripts denote partial derivatives, e.g.
$h_{\f\f}=\partial^2h/\partial \f^2$.

\ba \delta R_{tt} &=& - \o^2(h_{\f \f}+p_{\f \f}) +e^{2 (\n - \p)}
   h_{\f \f} +e^{2(\p-\n)} \o^4 p_{\f \f} +e^{2(\n-\a)} \Biggl\{ 2 \n_{rr}(h-k)
 \nonumber \\ 
   && + \ 2\n_r(h_r-k_r)+ 2\Bigl(\n_r+\r\Bigr)\n_r(h-k) +\frac{2}{r^2}
   \Bigl[\n_{\t\t}(h-p) +\n_\t(h_\t-p_\t) \nonumber \\
   && + \ \n_\t^2(h-p) \Bigr]+\n_r\Bigl[a_\f +2\p_r(h-k)+k_r+2p_r\Bigr] 
   + \frac{\n_\t}{r^2} \Bigl[2 \p_\t(h-p) \nonumber \\
   && + \ k_\t+2p_\t \Bigl]  +h_{rr}+ \frac{ h_{\t\t}}{r^2} + \Bigl(\p_r+\r 
   \Bigr)h_r +  \frac{\p_\t}{r^2} h_\t \Biggr\} \nonumber \\
   && + \ e^{2(\p-\a)} \Biggl\{ -\o^2\Bigl(p_{rr}+\frac{ p_{\t\t}}{r^2}
   \Bigr) +\o \p_r
    \Bigl[ 2\o \p_r(p-k) +\o a_\f -\o (h_r+k_r \nonumber \\
   &&+\ p_r) \Bigr] -\frac{ \o^2}{r^2}\p_\t(h_\t+k_\t+p_\t) 
   +\o\Bigl[ 2\o_r \p_r(k-p) +2\o \p_{rr}(k-p) \nonumber \\
   && +\ 2\o\p_r(k_r-p_r) \Bigr] +2\o^2\Bigl(\n_r+\r+2\p_r\Bigr)\p_r(k-p) 
  -\o^2 \n_r(2 a_\f+ p_r) \nonumber \\ 
   && -\ \frac{\o^2}{r^2}\n_\t p_\t -\frac{\o^2}{r} p_r\Biggr\}
   - \o^2e^{2(\p-\a-\n)} \Bigl[ \p_r \o L_\f +\frac{\p_\t}{r^2}\o M_\f \Bigr]
   \nonumber \\  
   && +\ e^{-2 \a} \Biggl\{ L_\f \Bigl[ \o_r +\o (2\p_r -\n_r) \Bigr] 
   +\frac{M_\f}{r^2}\Bigl[\o_\t+\o (2\p_\t-\n_\t)\Bigr] \Biggr\} \nonumber \\
   && +\ e^{4\p-2(\a+\n)} \Bigl[ 
   \o^3\o_r a_\f+\o^2\o_r^2(h+k-2p) +\sr\o^2\o_\t^2 (h-p)\Bigr]\nonumber \\
   && +\ e^{2(\p-\a)} \Biggl\{ (k-p)\Bigl[\o_r^2+\o\o_r \Bigl(4\p_r-2\n_r+
   \frac{2}{r}\Bigr)+2\o\o_{rr}\Bigr] \nonumber \\
   && +\ \o_r\o(h_r+k_r-4p_r) +\frac{\o_\t}{r^2}\o(h_\t-2p_\t-k_\t)\Biggr\} 
\ea

\ba   \delta R_{rr} &=& \Bigl(-e^{2(\a-\p)}+\o^2e^{2(\a-\n)} \Bigr) 
    k_{\f\f}-\frac{ k_{\t\t}}{r^2} -h_{rr}-2p_{rr}-a_{r\f}-\frac{k_\t}{r^2}
     (\n_\t+\p_\t) \nn \\
   && +\ k_r\Bigl(\n_r+\r+\p_r\Bigr) +\frac{2}{r^2} \Bigl[\a_{\t\t}(p-k)
   +\a_\t(p_\t-k_\t)  \nn \\
   && +\ \a_\t(\n_\t+\p_\t)(p-k) \Bigr] -\frac{\a_\t}{r^2}
   (h_\t+2p_\t-k_\t)
   +\a_r(h_r+k_r+a_\f) \nn \\
   &&-\ 2\p_r(p_r+a_\f)-2\n_r h_r-\frac{2}{r}p_r +\o e^{-2 \n} \Bigl(
   -\frac{\a_\t}{r^2} M_\f + \a_r L_\f -L_{r \f} \Bigr) \nonumber \\
   &&+\ \o_r e^{2(\p-\n)} \Bigl[-\o a_\f +\o_r(p-h)\Bigr]
\ea

\ba
   \delta R_{\t\t} &=& \Bigl( -e^{2(\a-\p)}+\o^2e^{2(\a-\n)}\Bigr)r^2p_{\f\f}
   -r^2p_{rr}-h_{\t\t}-k_{\t\t}-p_{\t\t}-r^2p_r\Bigl(\n_r+\r \nonumber \\
   && +\ \p_r\Bigr) +p_\t(\n_\t+\p_\t)+4r(\a_r+\r)(k-p)+2r^2\Bigl(\a_{rr}-\sr
   \Bigr)(k-p) \nonumber \\
   && +\ 2r^2\Bigl(\a_r+\r \Bigr)(k_r-p_r)+2\Bigl(\n_r-\r+\p_r\Bigr)r^2
    \Bigl(\a_r+\r\Bigr)(k-p) \nonumber \\
   && -r^2\Bigl(\a_r+\r\Bigr)(h_r+k_r+a_\f)+\a_\t(h_\t+2p_\t-k_\t)-2\p_\t 
   p_\t-2\n_\t h_\t \nn \\
   && +\ \o e^{-2\n}\Bigl[ -r^2\bigl(\a_r+\r\bigr)L_\f +\a_\t M_\f -M_{\t \f}
   \Bigr] + \o_\t^2e^{2(\p-\n)}(p-h)
\ea 

\ba
   \delta R_{tr} &=& -\frac{\o}{2}a_{\f\f}+\o(h_{r\f}-p_{r\f})+\o(2p_\f+h_\f
   -k_\f) (\n_r-\p_r)+\frac{\o^3}{2}e^{2(\p-\n)}a_{\f\f} \nn \\
   && +\ \oh e^{2(\p-\a)} \Biggl\{ -2a\o\p_r(\n_r+\r+\p_r)+\frac{\o}{r^2}
   \p_\t(-3a_\t-2a \p_\t-2a\n_\t) \nn \\
   && +\ \frac{\o}{r^2}(2\a_\t-\n_\t)a_\t-\frac{\o}{r^2} a_{\t\t}
   -2a\o\Bigl(\psi_{rr}
   +\frac{\psi_{\t\t}}{r^2}\Bigr) \Biggr\} +\oh\bigl(-e^{-2\p}+\o^2 e^{-2 \n}
   \bigr) L_{\f\f} \nn \\
   &&+\ \frac{\o_r}{2}(k_\f-h_\f-2p_\f) 
    - \oh a\o e^{4\p-2(\a+\n)}\Bigl(\o_r^2+\frac{\o_\t^2}{r^2}\Bigr) 
   +\frac{\o^2}{2}\o_r e^{2(\p-\n)}(h_\f+k_\f \nn \\
   &&-\ 4p_\f)  + \oh e^{2(\p-\a)}\Biggl\{a\o_r\Bigl(\n_r-3\p_r
   -\r\Bigr) 
   + \frac{ a}{r^2}\o_\t(\n_\t-3\p_\t)-a \Bigl(\psi_{rr}+\frac{\psi_{\t\t}}{r^2}
   \Bigr) \nn \\ 
   && -\ \frac{ \o_\t}{r^2} a_\t \Biggr\} 
   + \oh e^{2(\p-\a-\n)} \Biggl\{\o\o_r\Bigl[ L\Bigl(3\p_r-\n_r+\r\Bigr)
   +\frac{M}{r^2}(3\p_\t-2\a_\t-\n_\t) \nn \\
   && +\ \frac{M_\t}{r^2}\Bigl]+ \frac{\o}{r^2}\o_\t (2 L \a_\t -L_\t) 
   +\frac{\o_r}{r^2}\o_\t M
   +\o_r^2L+\sr\o\o_{r\t}M
   +\o\o_{rr}L \Biggr\} \nn \\
   && +\ e^{-2\a}\Biggl\{-L\Bigl(\frac{2}{r^2}\a_\t \n_\t+\r\n_r+\n_r^2+\n_r
   \p_r +\n_{rr}\Bigr) +\frac{M}{r^2}(2\a_\t\n_r-\n_r\n_\t  \nn \\ 
   && -\ \n_r \p_\t-\n_{r\t})+\frac{M_r}{r^2}\Bigl(\frac{\p_\t}{2}
   +\frac{\n_\t}{2}
   -\a_\t\Bigr) +\frac{L_\t}{r^2}\Bigl(-\frac{\p_\t}{2}+\frac{\n_\t}{2}+\a_\t
   \Bigr)-\frac{\n_r}{r^2}M_\t \nn \\
   &&-\ \frac{L_{\t\t}}{2r^2}+\frac{M_{r\t}}{2r^2}\Biggr\}
\ea

\ba 
   \delta R_{t \t} &=& \o(h_{\t\f}-p_{\t\f})+(\n_\t-\p_\t)\o(k_\f+h_\f)
   +\oh e^{2(\p-\a)} \Bigl[3\o\p_r a_\t - \o\Bigl(2\a_r+\r \nn \\
   && -\n_r\Bigr)a_\t+ \o a_{r\t}\Bigr] +\frac{1}{2}\Bigl(-e^{-2\p}
   +\o^2e^{-2\n}\Bigr)M_{\f\f}-\frac{\o_\t}{2}(h_\f+k_\f) \nn \\
   &&+\ \frac{ \o^2}{2}\o_\t 
   e^{2(\p-\n)}(h_\f-2p_\f-k_\f)
   +\oh e^{2(\p-\a)} \o_r a_\t \nn \\
   && +\ \oh e^{2(\p-\a-\n)} \Biggl\{ \o \o_\t \Bigl[L(3\p_r-2\a_r 
    - \n_r-\r)+\frac{M}{r^2}(3\p_\t-\n_\t) +L_r \Bigr]  \nn \\ 
   && +\ \o\o_r \Bigl[2M\Bigl(\a_r +\r\Bigr)-M_r\Bigr] + L\o_r\o_\t +L\o\o_{r
   \t} + \frac{M}{r^2}(\o_\t^2+\o\o_{\t\t})  \Biggr\} \nn \\
   &&  +\ e^{-2\a} \Biggl\{
   -\frac{M}{r^2}\Bigl[2r^2\n_r\Bigl(\a_r+\r\Bigr)+\n_\t^2+\n_\t\p_\t+\n_{\t\t}
   \Bigr] + L\Bigl[\Bigl(2\a_r+\r\Bigr)\n_\t \nn \\
   && -\ \n_r\n_\t-\n_\t\p_r-\n_{r\t} \Bigr]+\Bigl(\frac{\p_r}{2}+\frac{\n_r}
   {2}-\a_r+\frac{1}{2r}\Bigr)(L_\t-M_r)-L_r \n_\t \nn \\
   &&+\ \frac{L_{r\t}}{2}
   -\frac{M_{rr}}{2}    \Biggr\}   
\ea

\ba
   \delta R_{r\f} &=& -h_{r\f}-p_{r\f}+\p_r(h_\f+p_\f)+\n_r(k_\f-h_\f)
   +\Bigl(\a_r+\r\Bigr)(k_\f-p_\f) \nn \\
   && -\ \oh e^{2(\p-\n)}\o^2 a_{\f\f} +e^{2(\p-\a)} \Biggl\{ \frac{1}{2r^2}
   \Bigl[2a_\t\p_\t+2a\p_{\t\t}+a_{\t\t}+\n_\t(2a \p_\t+a_\t) \Bigr] \nn \\
   && +\ a \p_{rr}-\frac{\a_\t}{r^2} a_\t+a\p_r\Bigl(\n_r+\r+\p_r\Bigr)+
   \frac{\p_\t}{2r^2}(2a\p_\t+a_\t) \Biggr\} -\frac{\o}{2}e^{-2 \n}L_{\f\f}
   \nn \\
   && +\ \frac{ a}{2} e^{4\p-2(\a+\n)}\Bigl(\o_r^2+\frac{\o_\t^2}{r^2}\Bigr)
   +\frac{\o}{2}\o_r e^{2(\p-\n)}(4p_\f-k_\f-h_\f) \nn \\
   && +\oh e^{2(\p-\a-\n)} \Biggl\{\frac{\o_r}{2}(2\a_\t+\n_\t-3\p_\t)M 
   -\frac{2}{r^2}
   \omega_\t \a_\t L +L\o_r\Bigl(\n_r-3\p_r-\r\Bigr) \nn \\
   && +\frac{1}{r^2}\bigl(L_\t \o_\t -M_\t \o_r \bigr) 
   - L\o_{rr}-\frac{\o_{r\t}}{r^2}M \Biggr\}
\ea


\section{Angular Derivative Formulae}
\label{a:derivative}

In this appendix we derive high-order, finite-difference formulae that
approximate the angular derivatives of functions that are known at the
discrete set of angles $\mu_i=\cos\theta_i$ (with $i=1...n$), which
correspond to the zero's of the Legendre polynomial $P_{2n-1}(\mu_i) =0$.
A function $f(r,\mu )e^{im\phi}$ can be expanded in terms of associated
Legendre polynomials as

\be f(r,\mu )=\sum_{k=0}^\infty f_k(r)P^m_{k+|m|}(\mu ), \label{f1} 
\ee
Since $P^m_{k+|m|}$ is a polynomial of order $k$ multiplied by $(1-
\mu^2)^{|m|/2}$, expansion (\ref{f1}) is equivalent to
 
\be f(r,\mu ) = (1-\mu^2)^{|m|/2} \sum_{k=0}^\infty f_k^\dagger (r)P_k(\mu
), \label{f2} 
\ee
where $P_k(\mu )$ is the Legendre polynomial of order $k$.  By
orthogonality of the Legendre polynomials, the coefficients $f_k^\dagger
(r)$ in the expansion (\ref{f2}) are determined as
 
\be f_k^\dagger (r) = \left( k+{1\over 2}\right) \int_{-1}^1 f(r,\mu )(1-
\mu^2)^{-|m|/2} P_k(\mu )d\mu. \label{f3}
\ee
The integral in (\ref{f3}) can be computed with high accuracy using Gaussian
quadrature

\be \int_{-1}^1 g^+(r,\mu )d\mu \simeq \sum_{i=1}^n w_ig^+(r,\mu_i),
\label{f4} \ee
where $g^+(r,\mu )$ is a function symmetric in $\mu$ and $w_i$ are weights 
that are tabulated in e.g. Abramowitz \& Stegun \markcite{AbrSt}. For a 
function 
$g^-(r, \mu)$, antisymmetric in $\mu$, the integral in (\ref{f4}) vanishes. 
   
We are interested in functions $f(r,\mu )$ that have definite reflection
symmetry, i.e.\ that are of the form $f^\pm (r,\mu ) = \pm f^\pm (r,-\mu
)$.  For $f^+ (r,\mu )$, the nonvanishing coefficients in (\ref{f3})
become

\be f^\dagger_{2k} = \sum_{i=1}^n {1\over 2} (4k+1)w_i(1-\mu_i^2)^{-|m|/2}
P_{2k}(\mu_i)f^+ (r,\mu_i), \label{f5}
\ee
while for $f^-(r,\mu )$, (\ref{f3}) yields

\be f^\dagger_{2k+1} = \sum_{i=1}^n {1\over 2} (4k+3)w_i(1-\mu_i^2)^{-
|m|/2}P_{2k+1}(\mu_i)f^-(r,\mu_i ). \label{f6}
\ee
Differentiating (\ref{f2}) with respect to $\mu$ one obtains

\ba {\partial\over{\partial\mu}} f^+ (r,\mu_i) &=&
-{|m|\mu_i\over{(1-\mu^2_i)}} f^+ (r,\mu_i) \nonumber \\
& & + \ (1-\mu^2_i)^{|m|/2} \sum_{k=0}^\infty f_{2k}^\dagger (r)
{\partial\over{\partial\mu}} P_{2k}(\mu ), \label{f7}
\ea

\<and substituting (\ref{f5}) into (\ref{f7}) yields

\be {\partial\over{\partial\mu}} f^+ (r,\mu_i)=\sum_{j=1}^n
D^+_{ij}f^+ (r,\mu_j) \label{f8}, 
\ee 
where  

\ba D^+_{ij} &=& - {|m|\mu_i\over{(1-\mu_i^2)}} \delta_{ij}
+ (1-\mu_i^2)^{|m|/2} \nonumber \\ 
& & \sum_{k=0}^{n-1} {1\over 2} (4k+1)w_j(1-\mu_j^2)^{-|m|/2} P_{2k}(\mu_j)
{\partial\over{\partial\mu}} P_{2k}(\mu_i) \label{f9},
\ea

\<and $\delta_{ij}$ is the Kronecker delta. Next, we use the recurrence 
relation

\be {\partial\over{\partial\mu}} P_{2k}(\mu_i) = {2k\over{(1-\mu_i^2)}} \Bigl[-
\mu_iP_{2k}(\mu_i)+P_{2k-1}(\mu_i) \Bigr] \label{f10}, 
\ee
in (\ref{f9}) and define

\be S^+_{ij} = \sum_{k=1}^{n-1} k(4k+1)P_{2k}(\mu_j) \Bigl[-
\mu_iP_{2k}(\mu_i)+P_{2k-1}(\mu_i) \Bigr] .
\ee
In (\ref{f9}) we truncated the expansion to include Legendre polynomials
up to order $P_{2n-2}(\mu)$. 
If we repeat this for $f^-(r,\mu )$ and define

\be S^-_{ij} = \sum_{k=0}^{n-2} {(2k+1)\over 2} (4k+3)P_{2k+1}(\mu_j) 
\Bigl[-\mu_iP_{2k+1}(\mu_i)+P_{2k}(\mu_i) \Bigr], 
\ee
then $D^{\pm}_{ij}$ is given by
\be D^\pm_{ij} = {1\over{(1-\mu_i^2)}} \left\{ -|m| \mu_i\delta_{ij} +
\left[ {1-\mu_i^2\over{1-\mu_j^2}}\right]^{|m|/2} w_j S_{ij}^\pm\right\},
\ee
and the first-order angular derivative formula becomes
\be {\partial\over{\partial\mu}} f^\pm (r,\mu_i) = \sum_{j=1}^n D_{ij}^\pm
f^\pm (r,\mu_j).
\ee

In order to obtain the second-order angular derivatives, we differentiate
(\ref{f8}) with respect to $\mu$

\ba 
{\partial^2\over{\partial\mu^2}} f^+ (r,\mu_i) &= & -
\left[ {|m|(1+\mu_i^2)+m^2\mu_i^2\over{(1-\mu_i^2)^2}}\right] f^+
(r,\mu_i)
- {2|m|\mu_i\over{1-\mu_i^2}} {\partial\over{\partial\mu}} f^+(r,\mu_i) 
\nonumber \\
&& + \  \ (1-\mu_i)^{|m|/2} \sum_{k=0}^\infty f^\dagger_{2k}(r)
{\partial^2\over{\partial\mu^2}} P_{2k}(\mu_i). 
\ea
Differentiating the recurrence relation (\ref{f10}) with respect to $\mu$ we 
obtain

\ba \frac{\partial^2}{\partial\mu^2} P_{2k}(\mu_i) = \frac{2k}{ (1-
\mu_i^2)^2} & \Bigl\{ &  -  \Bigl[1+(1-2k)\mu_i^2 \Bigr] P_{2k}(\mu_i) 
\nonumber \\
& & + \ (3-4k)\mu_i P_{2k-1}(\mu_i) + (2k-1)P_{2k-2}(\mu_i) \  \Bigr\}.
\nonumber \\
&& 
\ea
Continuing in the same fashion as for the first-order derivatives, we define

\ba H_{ij}^\pm = {1\over{(1-\mu_i^2)^2}} &\Biggl\{& -
\Bigl[1+(1+|m|)\mu_i^2 \Bigr] |m|\delta_{ij} \nonumber \\
&&- 2|m| \mu_i(1-\mu_i^2)D_{ij}^\pm + 
\left[ {1-\mu_i^2\over{1-\mu_j^2}}\right]^{|m|/2}
w_j G_{ij}^\pm \ \Biggr\},
\ea
where

\ba G_{ij}^+ = \sum_{k=1}^{n-1} k(4k+1)P_{2k}(\mu_j) &\Bigl\{& -
\Bigl[1+(1-2k)\mu_i^2 \Bigr] P_{2k}(\mu_i) \nonumber \\
&&+ \ (3-4k)\mu_i P_{2k-1}(\mu_i)+ (2k-1)P_{2k-2}(\mu_i) \ \Bigr\}, 
\nonumber \\
&&
\ea
and 
\ba G_{ij}^- = \sum_{k=0}^{n-2} {(2k+1)\over 2}
(4k+3)P_{2k+1}(\mu_j) &\Bigl\{& - \Bigl(1-2k\mu_i^2 \Bigr)P_{2k+1}(\mu_i)
\nonumber \\
&&+ \ (1-4k)\mu_iP_{2k}(\mu_i)+ 2k P_{2k-1} \ \Bigr\}. \nonumber \\
&&
\ea

Then, the second-order angular derivatives for the functions $f^\pm (r,\mu)$ 
are \be {\partial^2\over{\partial\mu^2}} f^\pm (r,\mu_i) = \sum_{j=1}^n
H_{ij}^\pm f^\pm (r,\mu_j). 
\ee


\section{Perturbed Energy Conservation Equation}
\label{a:energy}

In this Appendix, we construct a variational principle for the perturbed
energy conservation equation (based on Friedman \& Ipser (1992)
\markcite{FriedmanIpser92}) and show its relation to the method presented
in section \ref{s:energy_eqn}. In addition, we derive an explicit expression 
for the inner product $<\delta U_j|L|\delta U_i>$ defined in section 
\ref{s:energy_eqn}.

\subsection{A Variational Principle}

In the Lagrangian formalism, perturbations are described by the Lagrangian 
displacement vector $\xi^a$. The complete description of a perturbation 
requires the solution of the perturbed field, Euler and energy conservation
equations for the metric perturbation $h_{ab}$ and the displacement vector
$\xi^a$. If one can solve all but one  of the above equations, using for
example a trial vector $\xi^a$, then the remaining equation will not be
satisfied. One then has to construct a criterion, which will be used to 
identify the equilibrium star for which the remaining equation is satisfied. 

For perturbations $(\xi^a,h_{ab})$ that satisfy the perturbed energy 
conservation equation

\be \delta \Bigl( u_b \nabla_a T^{ab} \Bigr) = 0, \label{energ}\ee

\<the perturbed field equations form a symmetric system (Friedman \& 
Schutz (1975) \markcite{F&S75b}). That is, two pairs  $(\xi^a,h_{ab})$ and 
$(\hat{\xi}^a,\hat{h}_{ab})$ satisfying (\ref{energ}) obey the symmetry 
relation

\be 
\hat{\xi}_b \ \delta \Bigl( \nabla_c T^{bc} \Bigr) +\frac{1}{16 \pi}
\hat{h}_{bc} \ \delta \Bigl( G^{bc} - 8 \pi T^{bc} \Bigr) 
= - {\cal{L}} \Bigr(\hat{\xi}^a,\hat{h}_{ab};\xi^a,h_{ab} \Bigl) 
  +\nabla_b \tilde{R}^b, \label{symmetric} \ee
  
\<where $G^{ab}=R^{ab}-\frac{1}{2}g^{ab}R_c{}^c$ and $\nabla_b \tilde{R}^b$ 
is a divergence constructed from $\xi^a$, $\hat{\xi}^a$, $h_{ab}$ and
$\hat{h}_{ab}$.
In (\ref{symmetric}), ${\cal{L}} \Bigr(\hat{\xi}^a,\hat{h}_{ab};\xi^a,h_{ab} 
\Bigl)$ is a function symmetric under the interchange of the trial solutions
$(\xi^a, h_{ab})$ and $(\hat{\xi}^a, \hat{h}_{ab})$; hence the r.h.s. in  
(\ref{symmetric}) is symmetric up to a divergence.

In the Eulerian approach to solving the perturbation equations, the 
perturbed Euler equation

\be \delta \Bigl(q^a{}_b \nabla_a T^{bc} \Bigr)=0, \label{AEuler}\ee
 
\<is solved analytically and six components of the perturbed field equation 
are solved
for $h_{ab}$ given a trial function $\delta U$. The symmetry in 
(\ref{symmetric}) can be exploited in order to construct a 
variational principle for the remaining unsolved equations. The solved 
perturbed Euler equation can be eliminated from (\ref{symmetric})
by decomposing the Lagrangian displacement vector $\xi^a$ into 
vectors normal and parallel to the 4-velocity

\be \xi^a= \xi^a_{\perp} - (\xi_cu^c)u^a, \label{xi}\ee

\<where $\xi^a_{\perp}=q^a{}_b \xi^b$. Equation (\ref{AEuler}) then implies
 
\be \hat{\xi}_b \ \delta \Bigl(\nabla_aT^{ab} \Bigr) = - (\xi_cu^c) \ \delta 
 \Bigl( u_b \nabla_a T^{ab} \Bigr). \ee
  
\<For trial solutions  the perturbed energy conservation 
equation is not satisfied  
 
\be \delta \Bigl( u_b \nabla_a T^{ab} \Bigr) \equiv L (\delta U) \neq 0, \ee

\<where $L$ is a linear operator acting on the function $\delta U$.

The Lagrangian displacement $\xi^a$ has a gauge freedom in its
component along $u^a$:  Adding a vector field $fu^a$ (where $f$ is some
arbitrary scalar function) to $\xi^a$ leaves the Eulerian perturbations
unchanged.  As was shown by Friedman \& Ipser (1992)
\markcite{FriedmanIpser92}, if (\ref{energ}) is not satisfied, the
perturbation equations can still be cast in a symmetric form as in
(\ref{symmetric}) (but with a redefined divergence term $\nabla_b
\tilde R^b$) if the component of $\xi^a$ along $u^a$ is given the value

\ba \xi_a u^a & = & (u \nabla)^{-1} \delta U \nonumber \\ 
               & = & \frac{1}{i \sigma u^t} \delta U. \ea

\<With this definition,

\be \hat{\xi}_b \ \delta \Bigl(\nabla_a T^{ab} \Bigr) = \frac{i \hat{\delta U}}
{\sigma u^t} L(\delta U). \ee

\<Next, we define

\ba {\cal F} (\hat{\delta U};\delta U) &\equiv& \frac{1}{16 \pi}
\hat{h}_{bc} \ \delta \Bigl( G^{bc} - 8 \pi T^{bc} \Bigr) \nonumber \\
 &=& \frac{1}{16 \pi} ( \hat{h}^{bc}-\frac{1}{2} g^{bc} \hat{h}_d{}^d)
 \ \delta \Bigl[R_{bc}-8\pi(T_{bc}-\frac{1}{2}g_{bc}T_d{}^d) \Bigr], \label{F}
\ea

\<which is implicitly bilinear in $\hat{\delta U}$ and $\delta U$. The gauge
freedom in $h_{ab}$ leaves only six (out of ten) components of the
perturbed field equations independent. Thus, if a
trial solution satisfies six components of the perturbed field equations 
in (\ref{F}),
the other four components of the perturbed field equations will be an 
implicit functional of the perturbed energy conservation equation, the only 
remaining equation that needs to be satisfied.
Schematically, one can write

\be {\cal F}(\hat{\delta U}; \delta U) \equiv \hat{\delta U} {\bf F}
 \Bigl[ L (\delta U) \Bigr], \ee

\<where ${\bf F}$ is a functional of $L (\delta U)$. We have used the linearity
of ${\cal F}(\hat{\delta U}; \delta U)$ to factor out $\hat{\delta U}$.
The symmetry relation (\ref{symmetric}) becomes
 
\ba -{\cal L} \Bigl( \hat{\xi}^a,\hat{h}_{ab};\xi^a,h_{ab} \Bigr) &=&
    \frac{i \hat{\delta U}}{\sigma u^t}L(\delta U)
   + \hat{\delta U} {\bf F}\Bigl[ L (\delta U) \Bigr] -\nabla_b \tilde{R}^b
  \nonumber \\
  &=& \hat{\delta U} \Bigl[ \frac{i}{\sigma u^t} + {\bf F} \Bigr] L 
     (\delta U) -\nabla_b \tilde{R}^b \nonumber \\
  &\equiv& - {\cal L}(\hat{\delta U}; \delta U)  
\ea

A variational principle for the perturbed energy conservation equation is
constructed, by requiring that the following integral (which is implicitly 
quadratic in $\delta U$) vanishes
  
\be I = \int -{\cal L} (\delta U; \delta U) \sqrt{-g} d^3x =0. \ee

\<Because the integral of the divergence vanishes, this yields
  
\be I= \int \delta U \Biggl[ \frac{i}{\sigma u^t} + {\bf F} \Biggr]
 L(\delta U) \sqrt{-g} d^3 x=0. \label{I}\ee
    
\<The integral in (\ref{I}) is stationary with respect to first order 
variations in $\delta U$
 
\ba & & \frac{\delta I}{\delta (\delta U)} =0, \label{dI} \\ 
 \Longrightarrow \ \ & &\Biggl[ \frac{i}{\sigma u^t} + {\bf F} \Biggr]
 L(\delta U) = 0, \\
 \Longrightarrow \ \  & & L(\delta U) =0, \label{L0}\ea

\<provided that $\frac{i}{\sigma u^t} L(\delta U) \neq -{\bf F}\Bigl[
L(\delta U) \Bigr]$. Thus, $I=0$ is a variational principle for the
perturbed energy conservation equation $L(\delta U)=0$. 

In practice, one can expand the trial function $\delta U$ in terms of
a set of basis functions $\delta U_i$

\be \delta U = \sum_i a_i \delta U_i. \label{dUsum}\ee

\<Substituting (\ref{dUsum}) into (\ref{I}) yields

\ba &&\sum_i \sum_j a_i a_j \int \delta U_j \Bigl[ \frac{i}{\sigma u^t}+
 {\bf F} \Bigr] L(\delta U_i) \sqrt{-g}d^3x=0 \nonumber \\
\Longrightarrow \ \ &&\sum_i \sum_j a_i a_j A_{(ij)}=0, \ea
 
\<where we defined the symmetric matrix $A$ with elements
 
\be A_{ij}=\int \delta U_j \Bigl[ \frac{i}{\sigma u^t}+
 {\bf F} \Bigr] L(\delta U_i) \sqrt{-g}d^3x.
\ee
 
\<Equation (\ref{dI}) implies that the integral I will be stationary to the 
variation of any of the coefficients $a_i$ in the expansion (\ref{dUsum}). 
Thus,

\be \frac{\delta I}{\delta a_i}=0 \ \ \Longrightarrow \ \ 
\sum_i a_j A_{(ij)}=0. \ee

\<The last equation is a homogeneous linear system for the coefficients $a_j$ 
which has a nontrivial solution only when 

\be {\rm det} A_{(ij)}=0. \label{Aij}\ee

Since we can find an explicit form for $A_{ij}$ in terms of known quantities,
we could have used (\ref{Aij}) as a criterion for locating neutral modes.
However, the term ${\bf F} [ L(\delta U_i)]$ involves many
second order angular and radial derivatives and it is not certain how
accurate its evaluation on our finite grid would be. Since the method described
in section \ref{s:energy_eqn} gives a substantially simpler 
criterion which does not involve ${\bf F}  [L(\delta U_i)]$, we chose to use 
that for locating the neutral modes. It is interesting to note that the
matrix elements

\be <\delta U_j |L| \delta U_i > = \int \frac{i \delta U_j}{\sigma u^t}
L(\delta U_i) \sqrt{-g} d^3x 
\ee
used in section \ref{s:energy_eqn} are {\it nearly}
symmetric under the interchange of $\delta U_i$ and $\delta U_j$, for all
configurations considered. This indicates that the method 
described in section \ref{s:energy_eqn} nearly coincides with a variational 
principle.

\subsection{An Expression for the Inner Product}

Since for the equilibrium star $u_b \nabla_a T^{ab}=0$,

\ba L(\delta U) &=& \delta \Bigl( u_b \nabla_a T^{ab} \Bigr) \nonumber \\
    &=&   \Delta \Bigl( u_b \nabla_a T^{ab} \Bigr) \nonumber \\
    &=& - \Delta \Bigl[ (\epsilon+P) \nabla_b u^b+u^b\nabla_b \epsilon
          \Bigr] \nonumber \\
    &=& -u^c \nabla_c \Bigl[ \Delta \epsilon + \frac{1}{2} (\epsilon+P)
         q^{ab} \Delta g_{ab} \Bigr] \nonumber \\
    &=& -i \sigma u^t \Bigl[ \Delta \epsilon + \frac{1}{2} (\epsilon+P)
         q^{ab} \Delta g_{ab} \Bigr], \label{LU1}
\ea

\<(cf. Friedman \& Ipser, 1992). Then,
 
\be L(\delta U) = -i \sigma u^t \Biggl\{ \delta \epsilon 
  + \xi^a_{\perp} \nabla_a \epsilon +\frac{1}{2} (\epsilon +P) 
    \Bigl[ h_c{}^c +u^a u^b h_{ab} +2 \nabla_a \xi^a +2u^au^b \nabla_a \xi_b
  \Bigr] \Biggr\},
\ee
\<where we used $\Delta g_{ab}=h_{ab}+\nabla_a \xi_b+\nabla_b \xi_a$. Using
the decomposition (\ref{xi}) of $\xi^a$, one obtains

\be \nabla_a \xi^a = \nabla_a\xi^a_{\perp} -\delta U, \ee

\<and

\be u^a u^b\nabla_a \xi_b =\delta U - \xi^b_{\perp}u^a \nabla_a u_b. \ee

\<The Euler equations for the equilibrium configuration yield

\be u^b \nabla_b u_c = -\frac{ q_c{}^b \nabla_bP}{\epsilon+P}. \ee

\<Then (\ref{LU1}) becomes

\be L(\delta U) = -i \sigma u^t \Biggl\{ \delta \epsilon 
  + \xi^a_{\perp} \nabla_a (\epsilon+P) +\frac{1}{2} (\epsilon +P) 
    \Bigl[ h_c{}^c +u^a u^b h_{ab} +2 \nabla_a \xi^a_{\perp}
  \Bigr] \Biggr\}.
\ee

\<The perturbed energy density is

\be \delta \epsilon = \frac{(\epsilon+P)^2}{P \Gamma} \Bigl( \delta U
 +\frac{1}{2} u^a u^b h_{ab} \Bigr), \ee

\<so that

\be L(\delta  U)= -i \sigma u^t \Biggl\{ (\epsilon+P) \Biggl[ 
\frac{(\epsilon+P)}{P \Gamma} \delta U +\frac{1}{2} \Biggl(1+ \frac{\epsilon+P}
{P \Gamma} \Biggr) u^a u^b h_{ab} +\frac{h_c{}^c}{2} \Biggr] 
 + \nabla_a \Bigl[\xi^a_{\perp}(\epsilon+P) \Bigr] \Biggr\}. \ee
 
\<The matrix elements defined in section \ref{s:energy_eqn} now take the form

\ba <\delta U_j|L|\delta U_i> &=& \int \frac{i \delta U_j}{\sigma u^t}
L(\delta U_i) \sqrt{-g} d^3x \nonumber \\
&=& \int(\epsilon+P) \Biggl\{ \delta U_j \Biggl[ \frac{(\epsilon+P)}{P \Gamma}
\delta U_i + \frac{1}{2} \Biggl( 1 + \frac{\epsilon+P}{P \Gamma} \Biggr)
 u^a u^b h_{ab} +\frac{h_c{}^c}{2} \Biggr] \nonumber \\ 
   && \quad\quad\quad\quad\quad- \  \xi^a_{\perp} \nabla_a \delta U_j \Biggr\} 
\sqrt{-g} d^3x, 
\ea
  
\<where we have used the time-independence of $\xi^a$ to eliminate the 
term $\int \nabla_a [ \xi^a_{\perp}
(\epsilon+P)\delta U] \sqrt{-g} d^3x$ as an integral of a spatial divergence.
 
Finally, the component of $\xi^a$ normal to the 4-velocity $u^a$ is related to 
the component of $\delta u^a$ normal to $u^a$ by

\be \xi^a_{\perp}= \frac{\delta u^a_{\perp}}{i \sigma u^t} =
 \frac{\delta u^a-\frac{1}{2}u^b u^c h_{bc}u^a}{i \sigma u^t}, \ee

\<(cf. Ipser \& Lindblom (1992) \markcite{IpserLindblom92}) and the 
expression for $<\delta U_j |L|\delta U_i>$ used in our numerical 
computations becomes

\ba <\delta U_j|L|\delta U_i> 
&=& \int(\epsilon+P) \Biggl\{ \delta U_j \Biggl[ \frac{(\epsilon+P)}{P \Gamma}
\delta U_i + \frac{1}{2} \Biggl( 1 + \frac{\epsilon+P}{P \Gamma}\Biggr)
 u^a u^b h_{ab} +\frac{h_c{}^c}{2} \Biggr] \nonumber \\
 && \quad\quad\quad\quad\quad - \ \frac{(\delta u^a-\frac{1}{2}u^b u^c 
  h_{bc}u^a)}{i \sigma u^t}\nabla_a \delta U_j \Biggr\} \sqrt{-g} d^3x, \ea
 
\<where $h_{ab}$ and $\delta u^a$ are computed with $\delta U_i$.


\clearpage
\begin{table}
\caption{Comparison of critical configurations in the full and truncated 
gauges} 

  \begin{center} 
  \begin{tabular}{ l c c } 
  \hline 
  \hline
   $N=1.0$  & &  \\
   $\bar{\epsilon}_c=0.3$  & $T/|W|_c$ & $\bar{\Omega}_c$   \\ 
   $7 \times 101$   & &  \\ 
 \hline
   \multicolumn{3}{c}{m=3}  \\ \hline
                        & &    \\        
    {\rm Full gauge} & 4.63e-2  & 2.82e-1 \\
    {\rm Truncated gauge} & 4.59e-2 & 2.81e-1 \\
                        & &    \\ \hline      
   \multicolumn{3}{c}{m=4}  \\ \hline
                        & &    \\        
    {\rm Full gauge} & 3.46e-2  & 2.48e-1 \\
    {\rm Truncated gauge} & 3.47e-2 & 2.48e-1 \\
                        & &    \\ \hline      
   \multicolumn{3}{c}{m=5}  \\ \hline
                        & &    \\        
    {\rm Full gauge} & 2.83e-2  & 2.26e-1 \\
    {\rm Truncated gauge} & 2.83e-2 & 2.26e-1 \\
                        & &    \\ \hline      
  \end{tabular}
  \end{center}
\label{t:trunc} 
\end{table}

\clearpage
\begin{table}
 
\caption{Comparison of critical $T/|W|$ in Newtonian limit with other authors}
  
  \begin{center} 
  \begin{tabular}{ l r r r } 
  \hline 
  \hline
     & & & \\
     & m=3 & m=4 & m=5  \\ 
     & & & \\ 
 \hline
   \multicolumn{4}{c}{N=1.0}  \\ \hline
                       & & &    \\        
    {\rm Present}                              & 7.92e-2 & 5.79e-2 & 4.62e-2 \\
    {\rm Managan (1985) \markcite{Managan85}}  & 7.94e-2 & 5.81e-2 &  -----  \\
    {\rm Imamura {\it et al.} (1985) \markcite{IFD85}}    & 8.0e-2 & 5.8e-2 
    & 4.4e-2 \\
    {\rm Ipser \& Lindblom (1990) \markcite{IpserLindblom90}} & 8.00e-2 
    & 5.84e-2 & 4.53e-2 \\ 
                       & & &    \\ \hline      
   \multicolumn{4}{c}{N=1.5}  \\ \hline
                       & & &    \\        
    {\rm Present}                              & 5.61e-2 & 4.33e-2 & 3.36e-2 \\
    {\rm Managan (1985)\markcite{Managan85}}   & 5.6e-2 & 4.3e-2 &  -----    \\
    {\rm Imamura {\it et al.} (1985) \markcite{IFD85}}    & 5.7e-2 & 4.3e-2 
    &  ----- \\
                       & & &    \\ \hline      
   \multicolumn{4}{c}{N=2.0}  \\ \hline
                       & & &    \\        
    {\rm Present}                              & 3.35e-2 & 2.81e-2 & 2.28e-2 \\
    {\rm Managan (1985) \markcite{Managan85}}             & 3.3e-2 & 2.8e-2 &
      -----    \\
                        & & &    \\ \hline      
  \end{tabular}
  \end{center}
\label{t:Newtonian} 
\end{table}

\clearpage

\begin{table} 

\caption{Critical configurations for N=1.0 polytropes}

  \begin{center} 
  \begin{tabular}{ l c c c c c} 
  \hline 
  \hline
     & & & & & \\ 
     & $\bar{\epsilon}_c$ & $T/|W|_c$ &  $(T/|W|_c)/$ & $\bar{\Omega}_c$ & 
                                                    $\Omega_c/ \Omega_K$ \\
     &            &         &  $(T/|W|_K)$  &          &    \\
     & & & & & \\ 
  \hline
  & \multicolumn{4}{c}{m=2} & \\ \hline
                       & & & & &   \\        
    {\rm Relativistic} & 3.4e-1 & 6.49e-2 & 0.777 & 3.44e-1 & 0.911 \\
                       & & & & &  \\ \hline  
  & \multicolumn{4}{c}{m=3} & \\ \hline
                       & & & & &   \\        
    {\rm Newtonian}    & 1.0e-8 & 7.92e-2 & 0.769 & 6.69e-5 & 0.921 \\
    {\rm Relativistic} & 3.4e-1 & 4.55e-2 & 0.544 & 2.96e-1 & 0.783 \\
                       & & & & &  \\ \hline  
  & \multicolumn{4}{c}{m=4} & \\ \hline
                       & & & & &   \\        
    {\rm Newtonian}    &  1.0e-8 & 5.79e-2 & 0.562 & 5.94e-5 & 0.818 \\
    {\rm Relativistic} & 3.4e-1 & 3.53e-2 & 0.422 & 2.64e-1 & 0.699 \\
                       & & & & &  \\ \hline
  & \multicolumn{4}{c}{m=5} & \\ \hline
                       & & & & &   \\        
    {\rm Newtonian}    & 1.0e-8 & 4.62e-2 & 0.449 & 5.41e-5 & 0.745 \\
    {\rm Relativistic} & 3.4e-1 & 2.87e-2 & 0.343 & 2.40e-1 & 0.635 \\
                       & & & & &  \\ \hline 
  \end{tabular}
  \end{center}
\label{t:1.0} 
\end{table}

\clearpage
\begin{table}
 
\caption{Critical configurations for N=1.5 polytropes}
  
  \begin{center} 
  \begin{tabular}{ l c c c c c} 
  \hline 
  \hline
     & & & & & \\ 
     & $\bar{\epsilon}_c$ & $T/|W|_c$ &  $(T/|W|_c)/$ & $\bar{\Omega}_c$ & 
                                                    $\Omega_c/ \Omega_K$ \\
     &            &         &  $(T/|W|_K)$  &          &    \\
     & & & & & \\ 
  \hline
  & \multicolumn{4}{c}{m=3} & \\ \hline
                       & & & & &   \\        
    {\rm Newtonian}    & 1.0e-7 & 5.61e-2 & 0.943 & 1.62e-4 & 0.980 \\
    {\rm Relativistic} & 6.1e-2 & 3.82e-2 & 0.804 & 1.02e-1 & 0.917 \\
                       & & & & &  \\ \hline  
  & \multicolumn{4}{c}{m=4} & \\ \hline
                       & & & & &   \\        
    {\rm Newtonian}    & 1.0e-7 & 4.33e-2 & 0.728 & 1.47e-4 & 0.886 \\
    {\rm Relativistic} & 6.1e-2 & 2.79e-2 & 0.587 & 8.87e-2 & 0.800 \\
                       & & & & &  \\ \hline
  & \multicolumn{4}{c}{m=5} & \\ \hline
                       & & & & &   \\         
    {\rm Newtonian}    & 1.0e-7 & 3.36e-2 & 0.565 & 1.32e-4 & 0.796 \\
    {\rm Relativistic} & 6.1e-2 & 2.34e-2 & 0.492 & 8.18e-2 & 0.738 \\
                       & & & & &  \\ \hline 
  \end{tabular}
  \end{center}
\label{t:1.5} 
\end{table}

\clearpage
\begin{table}
 
\caption{Critical configurations for N=2.0 polytropes}
  
  \begin{center} 
  \begin{tabular}{ l c c c c c} 
  \hline 
  \hline
     & & & & & \\ 
     & $\bar{\epsilon}_c$ & $T/|W|_c$ &  $(T/|W|_c)/$ & $\bar{\Omega}_c$ & 
                                                    $\Omega_c/ \Omega_K$ \\
     &            &         &  $(T/|W|_K)$  &          &    \\
     & & & & & \\ 
  \hline
  & \multicolumn{4}{c}{m=3} & \\ \hline
                       & & & & &   \\        
    {\rm Newtonian}    & 1.0e-8 & 3.35e-2 & 0.991 & 3.67e-5 & 0.997 \\
    {\rm Relativistic} & 5.1e-3 & 2.59e-2 & 0.951 & 2.22e-2 & 0.979 \\
                       & & & & &  \\ \hline  
  & \multicolumn{4}{c}{m=4} & \\ \hline
                       & & & & &   \\        
    {\rm Newtonian}    & 1.0e-8 & 2.81e-2 & 0.832 & 3.42e-5 & 0.928 \\
    {\rm Relativistic} & 5.1e-3 & 2.10e-2 & 0.771 & 2.02e-2 & 0.895 \\
                       & & & & &  \\ \hline
  & \multicolumn{4}{c}{m=5} & \\ \hline
                       & & & & &   \\        
    {\rm Newtonian}    & 1.0e-8 & 2.28e-2 & 0.676 & 3.12e-5 & 0.848 \\
    {\rm Relativistic} & 5.1e-3 & 1.70e-2 & 0.624 & 1.84e-2 & 0.814 \\
                       & & & & &  \\ \hline 
  \end{tabular}
  \end{center}
\label{t:2.0} 
\end{table}


\newpage
 
\figcaption[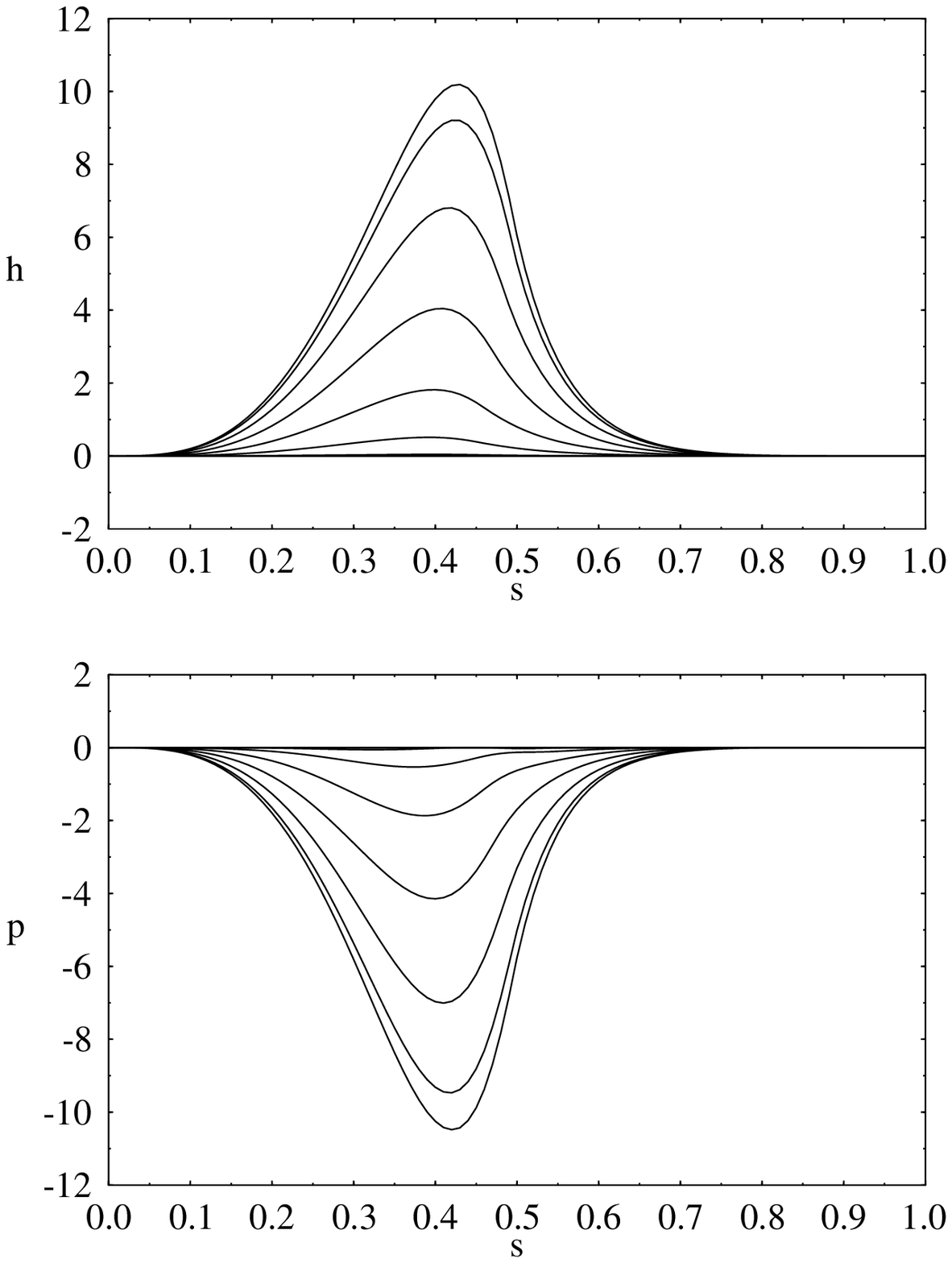]{Representative solution for the perturbation
functions $h$ and $p$ (at $\phi=0$). Notice that $p \simeq -h$ even for
this very relativistic configuration.  The scaling of the vertical axis
is determined by the trial function $\delta U$.  The equator of the
star is at $s=0.5$, while $s=1.0$ corresponds to infinity.  The
solution in each figure is shown at $\mu \equiv \cos \theta = 0$, 0.23,
0.45, 0.64, 0.80, 0.92, 0.98 and 1.0. The maximum is at $\mu=0$.
\label{fig1}}

\figcaption[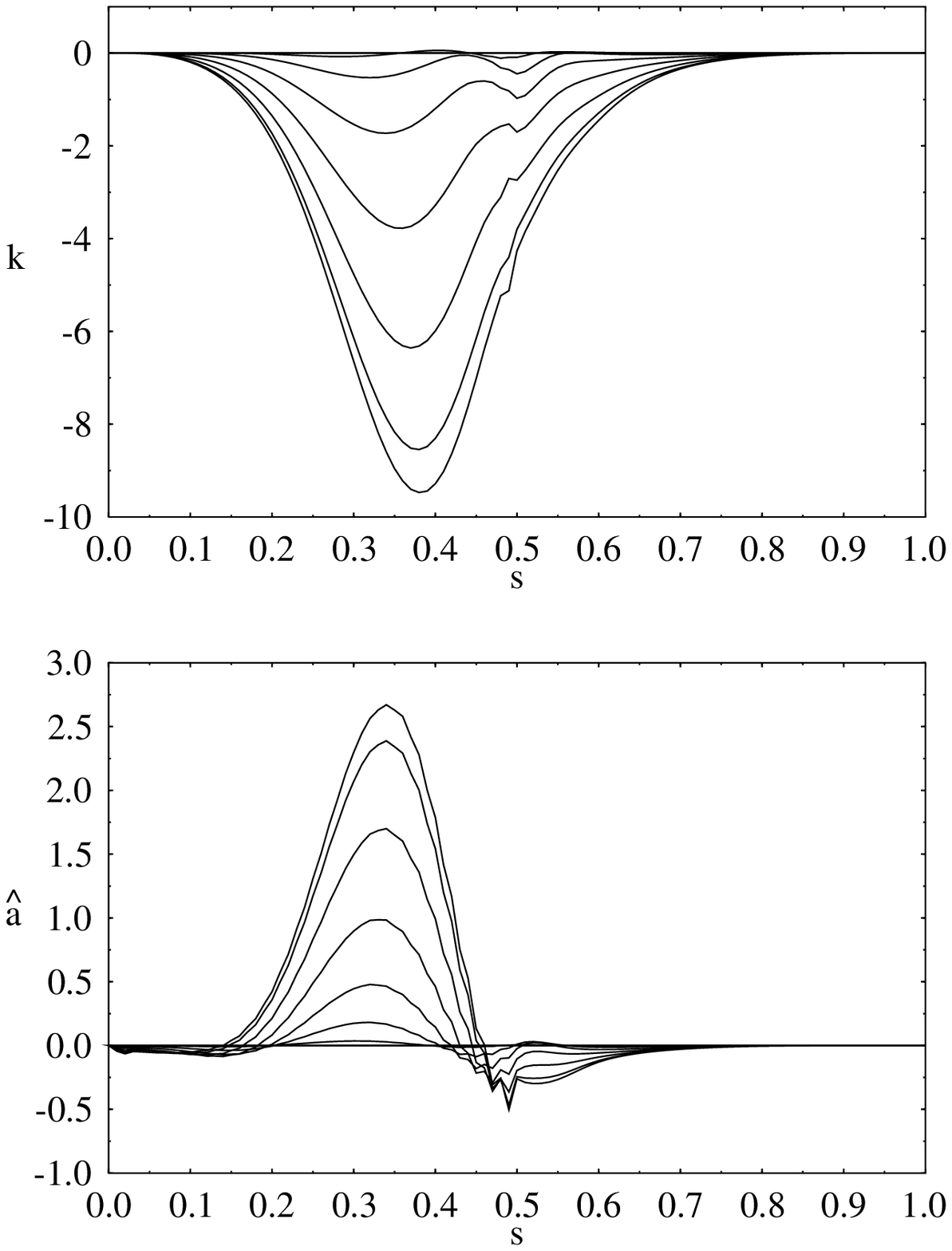]{Representative solution for the perturbation 
functions $k$ (at $\phi=0$) and 
$\hat a$ .
Notice that $k \simeq -h$.
The scaling of the vertical axis is determined by the trial function $\delta 
U$.
The equator of the star is at $s=0.5$, while $s=1.0$ corresponds to infinity.
The solution in each figure is shown at $\mu \equiv \cos \theta = 0$,
0.23, 0.45, 0.64, 0.80, 0.92, 0.98 and 1.0. The maximum is at $\mu=0$.
\label{fig2}}

\figcaption[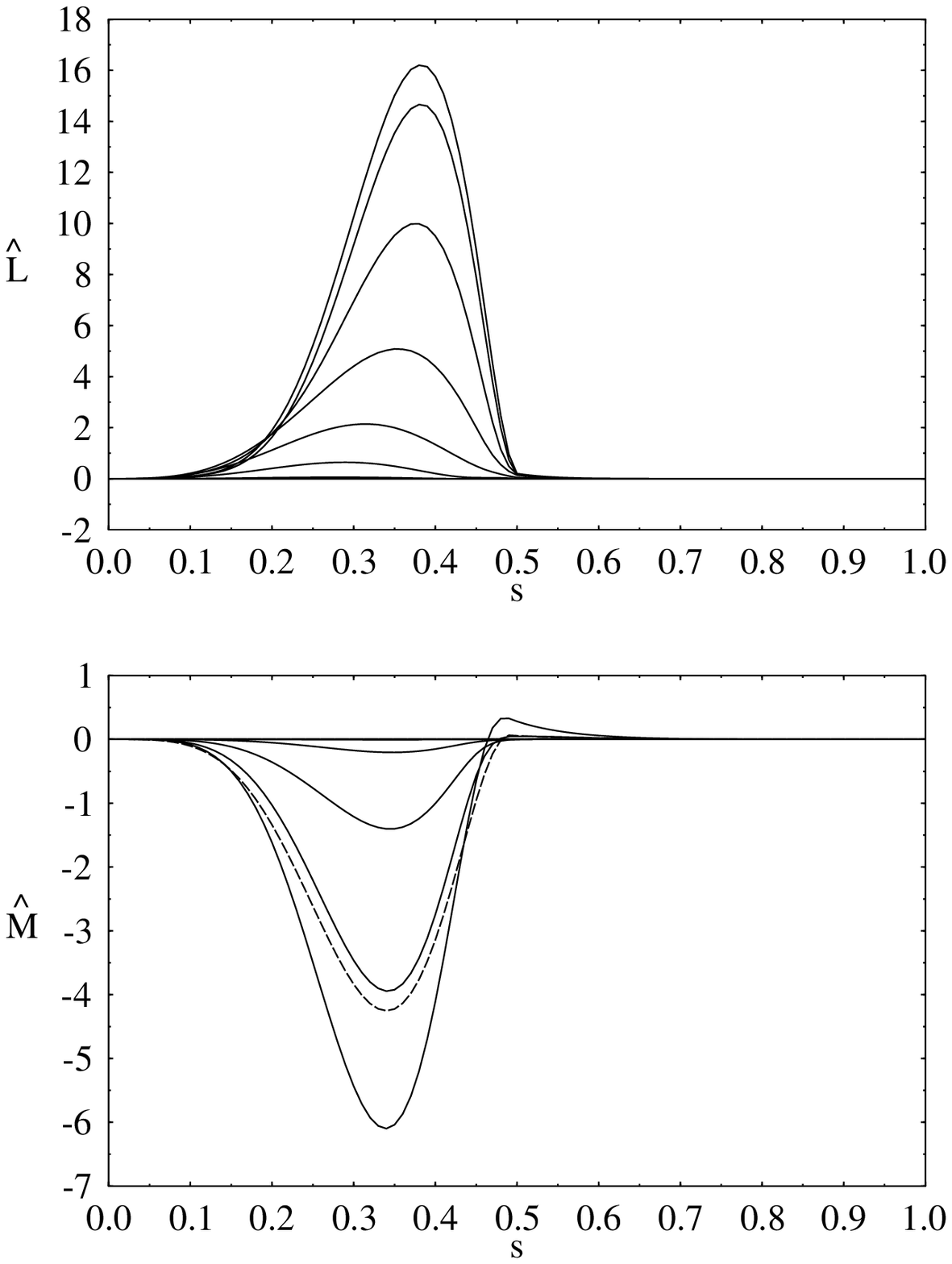]{Representative solution for the perturbation 
functions $\hat L$ and $\hat M$.
The scaling of the vertical axis is determined by the trial function $\delta 
U$.
The equator of the star is at $s=0.5$, while $s=1.0$ corresponds to infinity.
The solution in each figure is shown at $\mu \equiv \cos \theta = 0$,
0.23, 0.45, 0.64, 0.80, 0.92, 0.98 and 1.0. For $\hat L$ the maximum is at 
$\mu=0$. For $\hat M$ the dashed line is at $\mu=0.23$ and the maximum is 
at $\mu=0.45$.
\label{fig3}}

\figcaption[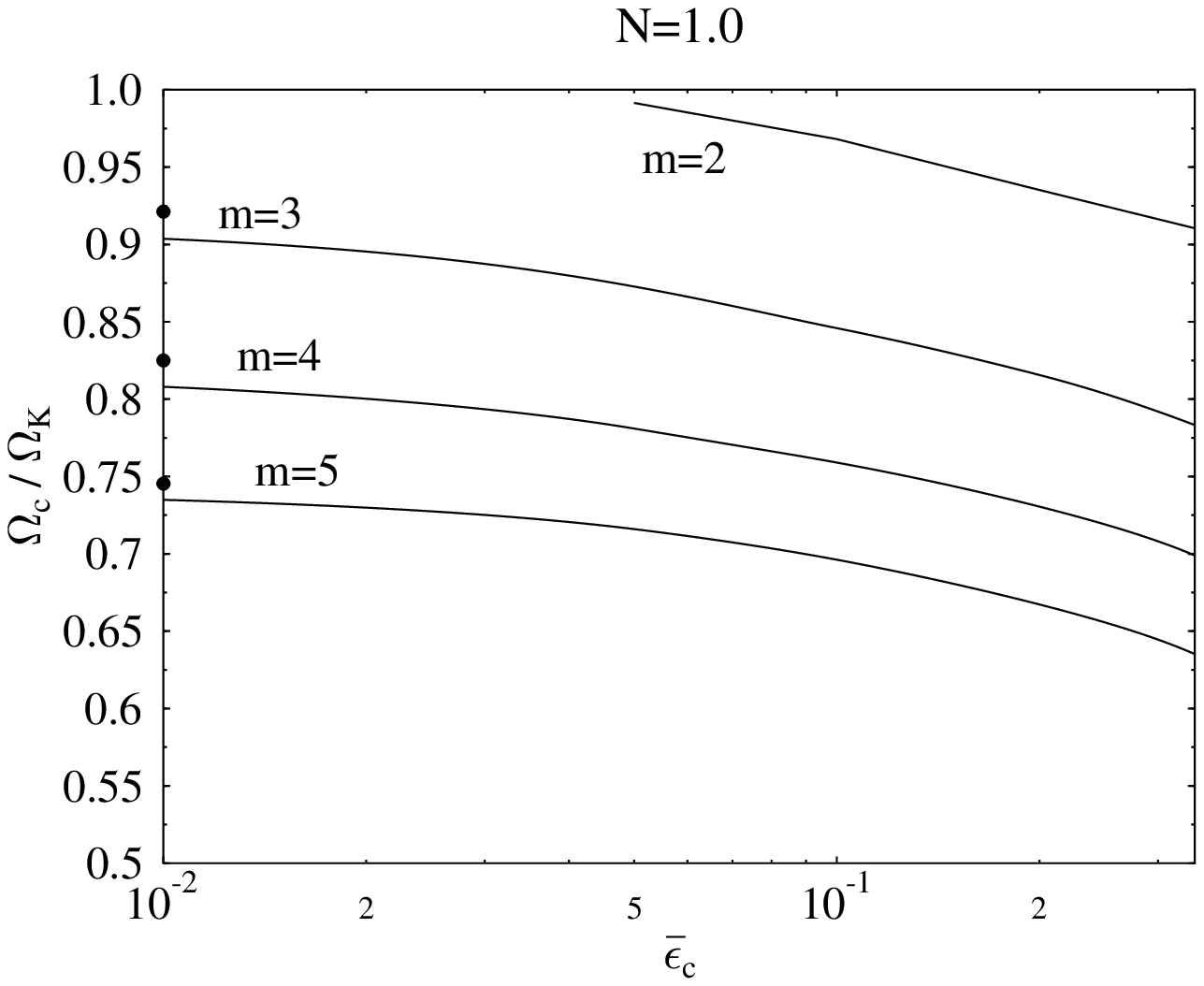]{Critical angular velocity over Keplerian angular 
velocity at same
central energy density vs. the dimensionless central energy density 
$ \bar{\epsilon}_c$ for the $m=2$, 3, 4 and 5 neutral modes of $N=1.0$ 
polytropes. 
The largest value of $ \bar{\epsilon}_c$ shown corresponds to the most 
relativistic stable configurations, while the lowest $ \bar{\epsilon}_c$ 
corresponds to less relativistic configurations. The filled circles on the 
vertical axis represent the Newtonian limit. \label{fig4}}

\figcaption[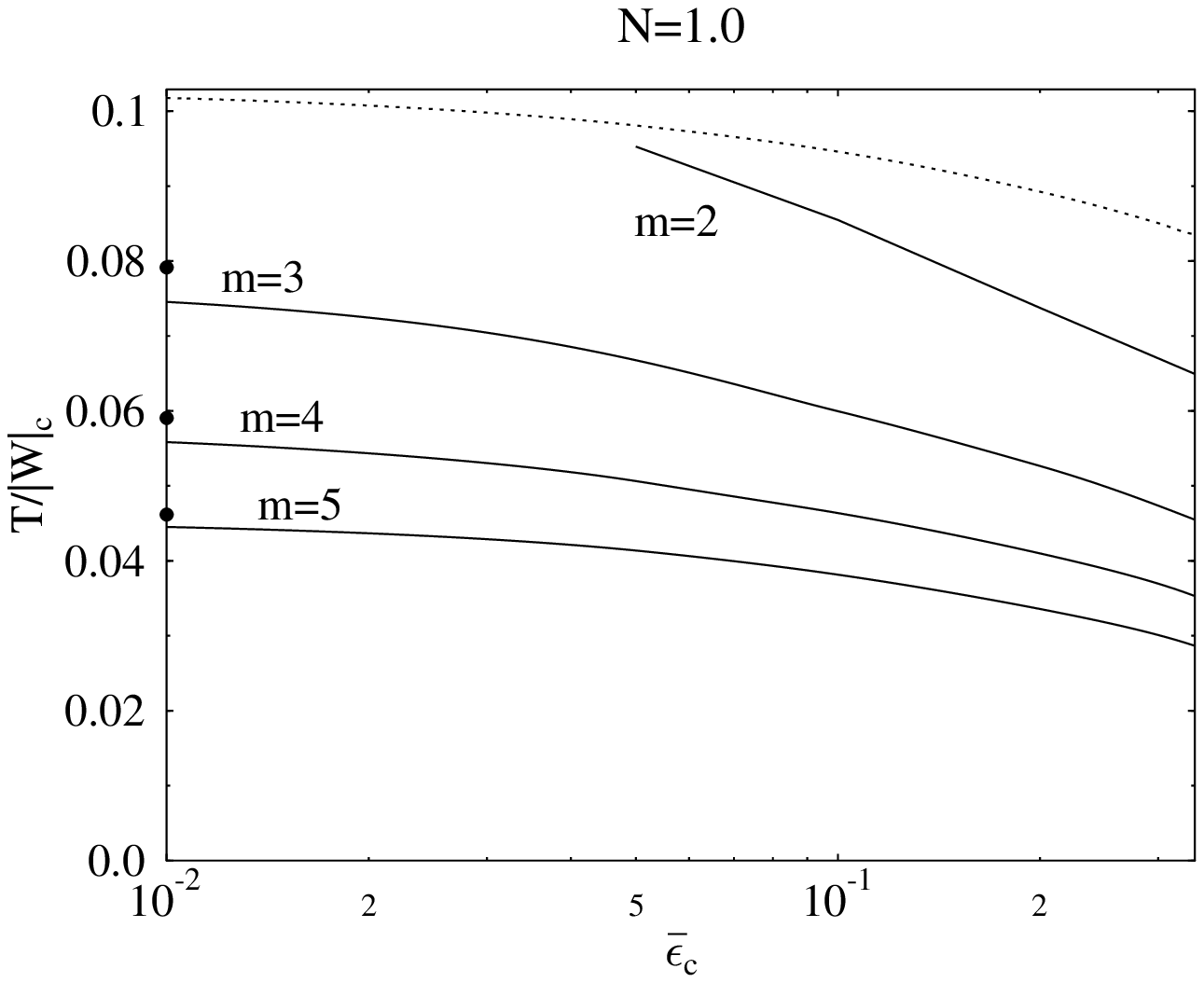]{Critical ratio of rotational to gravitational binding 
energy
vs. the dimensionless central energy density 
$ \bar{\epsilon}_c$ for the $m=2$, 3, 4 and 5 neutral modes of 
$N=1.0$ polytropes. 
The largest value of $ \bar{\epsilon}_c$ shown corresponds to the most 
relativistic stable configurations, while the lowest $ \bar{\epsilon}_c$ 
corresponds to less relativistic configurations. The filled circles on the 
vertical axis represent the Newtonian limit while the dotted line is the Kepler
limit. \label{fig5}}

\figcaption[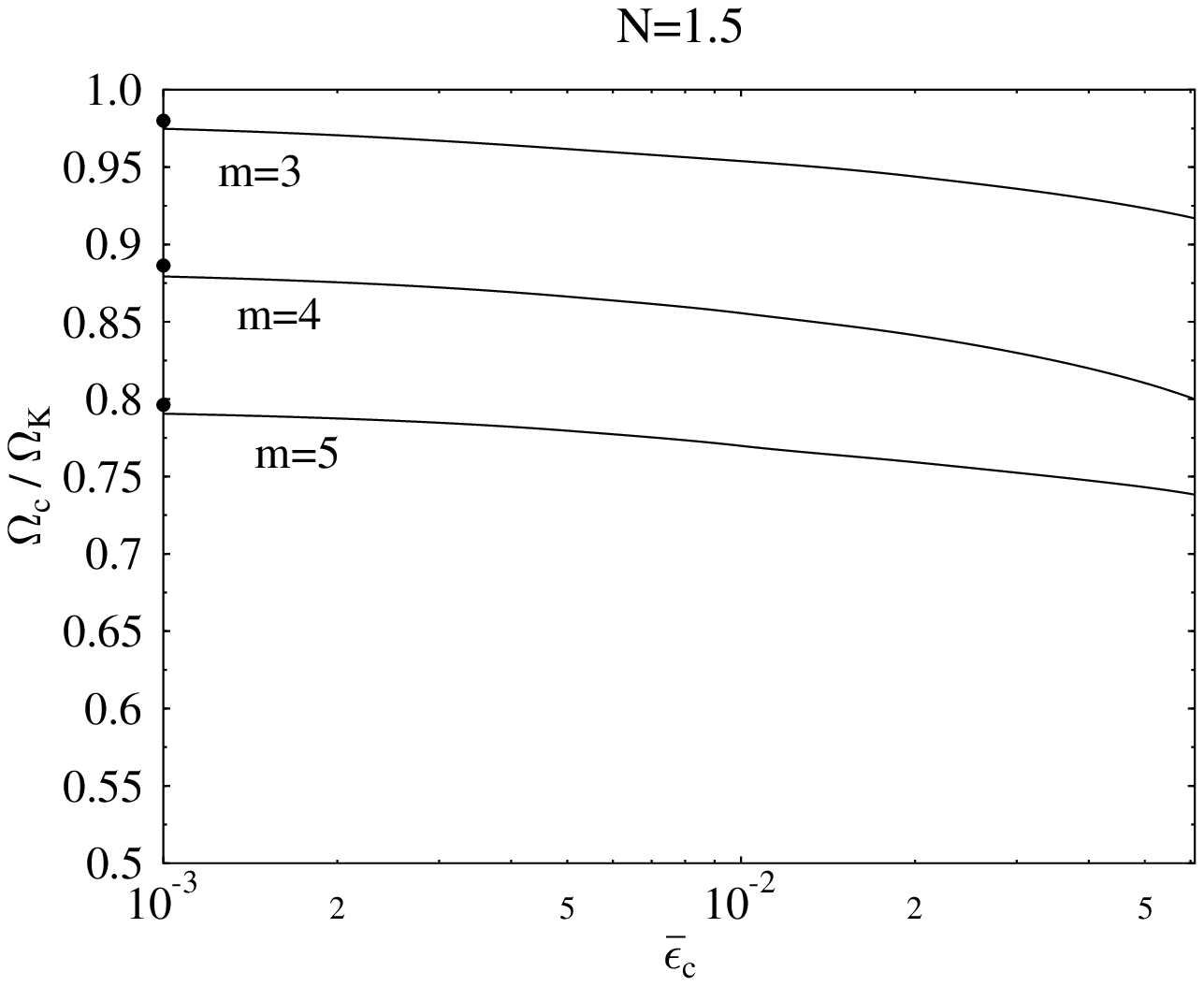]{Critical angular velocity over Keplerian angular 
velocity at same
central energy density vs. the dimensionless central energy density 
$ \bar{\epsilon}_c$ for the $m=3$, 4 and 5 neutral modes of $N=1.5$ 
polytropes. 
The largest value of $ \bar{\epsilon}_c$ shown corresponds to the most 
relativistic stable configurations, while the lowest $ \bar{\epsilon}_c$ 
corresponds to less relativistic configurations. The filled circles on the 
vertical axis represent the Newtonian limit. \label{fig6}}

\figcaption[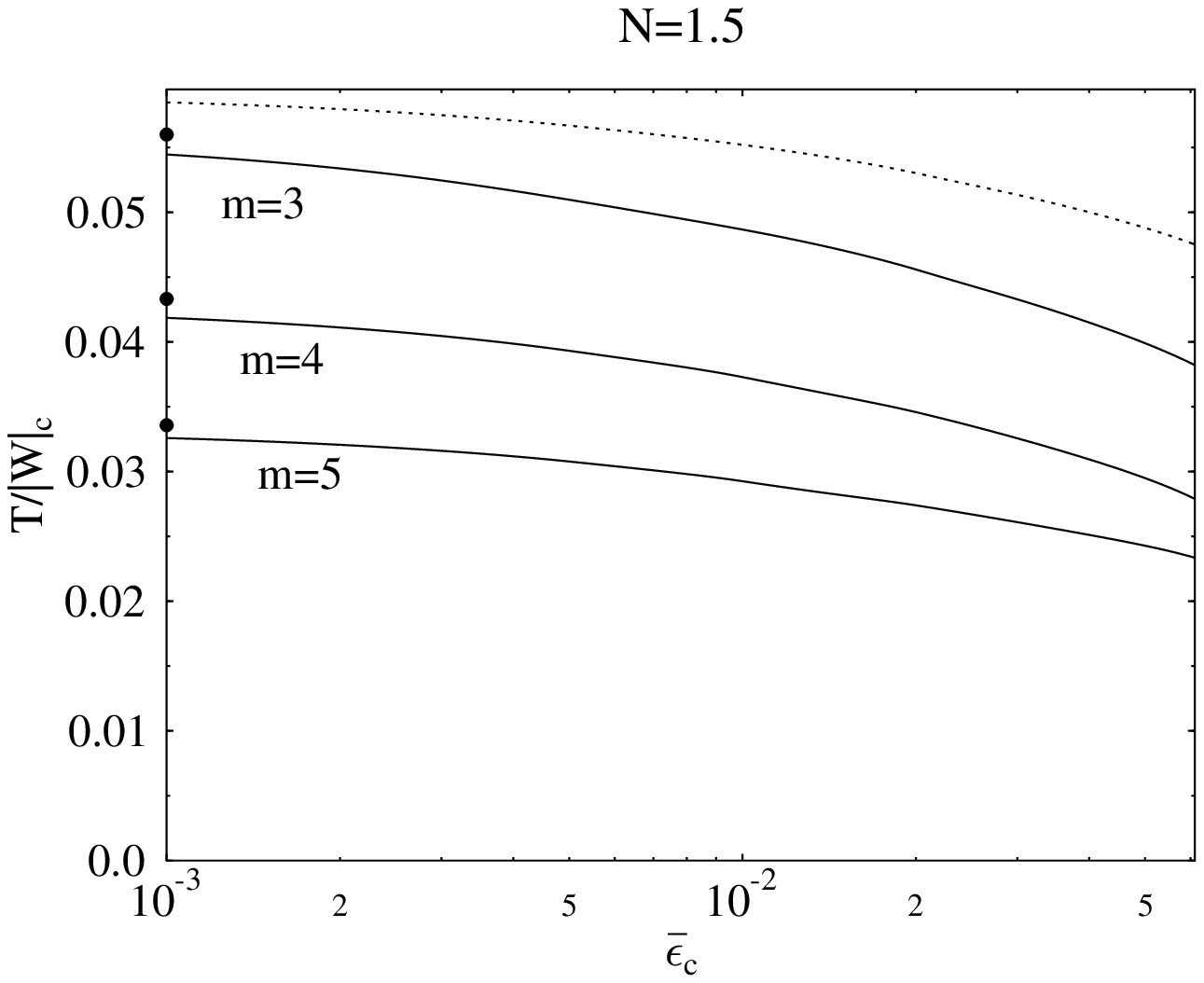]{Critical ratio of rotational to gravitational binding 
energy
vs. the dimensionless central energy density 
$ \bar{\epsilon}_c$ for the $m=3$, 4 and 5 neutral modes of $N=1.5$ 
polytropes. 
The largest value of $ \bar{\epsilon}_c$ shown corresponds to the most 
relativistic stable configurations, while the lowest $ \bar{\epsilon}_c$ 
corresponds to less relativistic configurations. The filled circles on the 
vertical axis represent the Newtonian limit  while the dotted line is the 
Kepler limit. \label{fig7}}

\figcaption[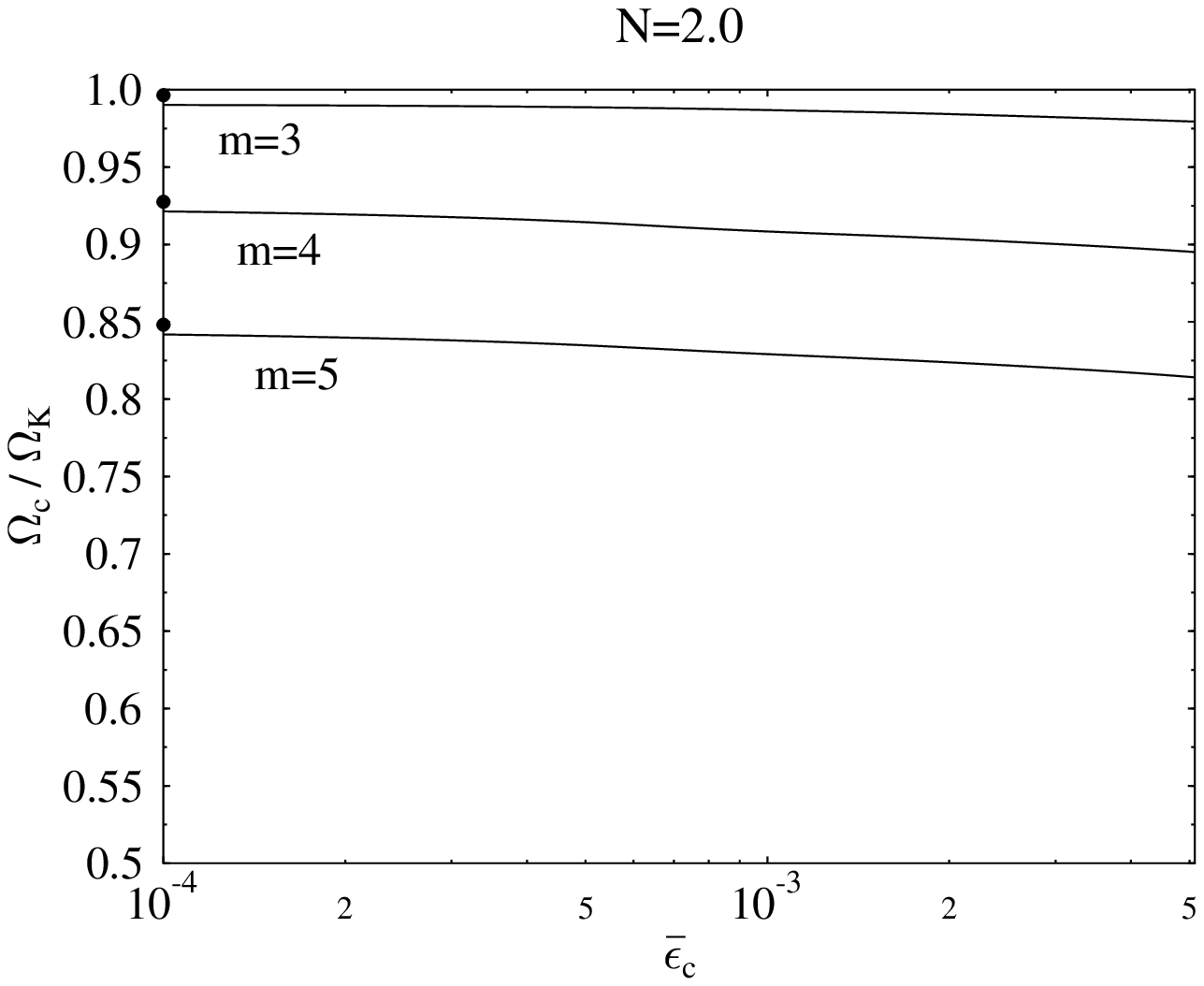]{Critical angular velocity over Keplerian angular 
velocity at same
central energy density vs. the dimensionless central energy density 
$ \bar{\epsilon}_c$ for the $m=3$, 4 and 5 neutral modes of $N=2.0$ 
polytropes. 
The largest value of $ \bar{\epsilon}_c$ shown corresponds to the most 
relativistic stable configurations, while the lowest $ \bar{\epsilon}_c$ 
corresponds to less relativistic configurations. The filled circles on the 
vertical axis represent the Newtonian limit. \label{fig8}}

\figcaption[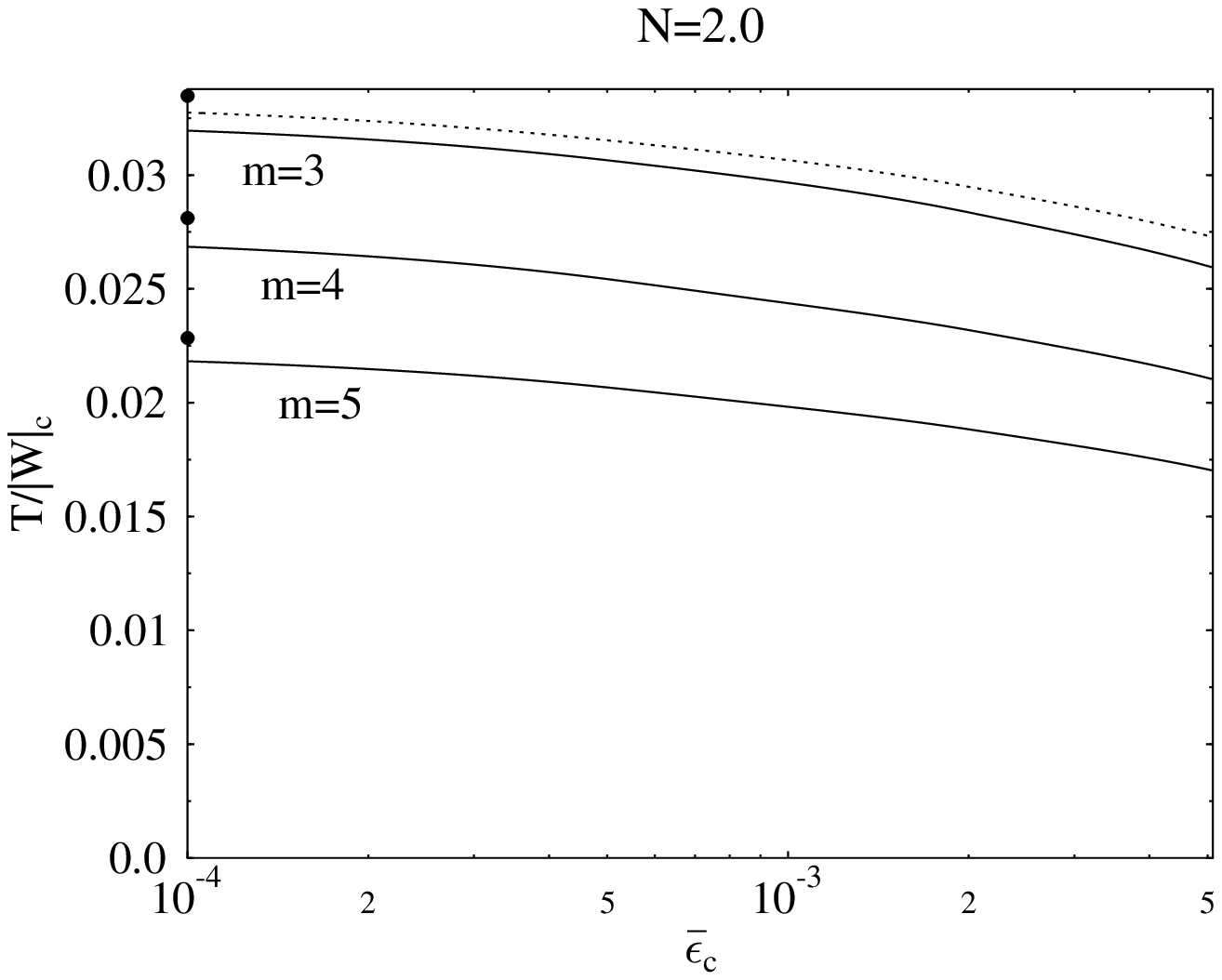]{Critical ratio of rotational to gravitational binding 
energy
vs. the dimensionless central energy density 
$ \bar{\epsilon}_c$ for the $m=3$, 4 and 5 neutral modes of $N=2.0$ 
polytropes. 
The largest value of $ \bar{\epsilon}_c$ shown corresponds to the most 
relativistic stable configurations, while the lowest $ \bar{\epsilon}_c$ 
corresponds to less relativistic configurations. The filled circles on the 
vertical axis represent the Newtonian limit while the dotted line is the Kepler
limit. \label{fig9}}


\begin{references}

\reference{Andersson97} Andersson, N. 1997, to appear.

\reference{AbrSt} Abramowitz, M. \& Stegun, I. A., 
  {\it Handbook of Mathematical Functions}, (New York, Dover, 1964)


\reference{BFG96} Bonazzola, S., 
   Frieben, J. \& Gourgoulhon, E. 1996, ApJ, {\bf 460}, 379

\reference{Chandra70} Chandrasekhar, S. 1970, Phys. 
   Rev. Lett., {\bf 24}, 611

\reference{CST94} Cook, G. B., Shapiro, S. L., 
   \& Teukolsky, S. A. 1994, ApJ, {\bf 424}, 823

\reference{Cutler91} Cutler, C. 1991, ApJ, {\bf 374}, 
   248

\reference{CutlerLindblom92} Cutler, C. \& Lindblom, L.
   1992, ApJ, {\bf 385}, 630

\reference{comparison} Eriguchi Y., Gourgoulhon E., Nozawa T.,
Stergioulas N. 1996 (names in temporary order, in preparation).

\reference{FriedmanIpser92} Friedman, J. L., \& Ipser, 
   J. R. 1992, Phil. Trans. R. Soc. Lond., {\bf A340}, 391

\reference{F&M97} Friedman, J. L. \& Morsink, S. 1997, to appear.

\reference{F&S75b} Friedman, J. L. \& Schutz, B. F. 
   1975, ApJ, {\bf 200}, 204 

\reference{F&S78a} Friedman, J. L  \& Schutz, B. F. 
   1978a, ApJ, {\bf 221}, 937 

\reference{F&S78b} Friedman, J. L. \& Schutz, B. F. 
   1978b, ApJ, {\bf 222}, 281

\reference{GroppSmith94} Gropp, W. \& Smith, B., in {\it 
   Proceedings of the Scalable Parallel Libraries Conference} (IEEE,1994) 
   pg. 60 

\reference{IFD85} 
   Imamura, J. N., Friedman, J. L. \& Durisen, R. H. 1985,
   ApJ, {\bf 294}, 474

\reference{IpserLindblom90} Ipser, J. R. \& Lindblom, L.
   1990, ApJ, {\bf 355}, 226

\reference{IpserLindblom91a} Ipser, J. R. \& Lindblom, 
   L. 1991a, ApJ, {\bf 373}, 213

\reference{IpserLindblom91b} Ipser, J. R. \& Lindblom, 
   L.  1991b, ApJ, {\bf 379}, 285

\reference{IpserLindblom92} Ipser, J. R. \& Lindblom, L.
   1992, ApJ, {\bf 389}, 392

\reference{IpserManagan85} Ipser, J. R. \& Managan, R. A.
   1985, ApJ, {\bf 292}, 517

\reference{James64} James, R.A. 1964, ApJ, {\bf 140}, 
   552

\reference{KEH89b} Komatsu, H., Eriguchi, 
   Y., \& Hachisu, 1989, \mnras, {\bf 239}, 153 

\reference{Lindblom95} Lindblom, L. 1995, ApJ, 
   {\bf 438}, 265

\reference{LM95} Lindblom, L. \& Mendell, G. 1995, ApJ, {\bf 444}, 804.

\reference{Lynden-BellOstriker67} Lynden-Bell, D. 
   \& Ostriker, J. P. 1967, \mnras, {\bf 136}, 293

\reference{Managan85} Managan, R. A. 1985, ApJ, 
   {\bf 294}, 463

\reference{Priou92} Priou, D. 1992, \mnras, {\bf 254}, 435

\reference{ReggeWheeler57} Regge, T. \& Wheeler, J. A. 
   1957, Phys. Rev. {\bf 108}, 1063

\reference{SkinnerLindblom96} Skinner, D. \& 
   Lindblom, L. 1996, preprint, Montana State University

\reference{S96} Stergioulas, N. 1996, ``Structure and Stability of Rotating Relativistic Stars,'' Ph.D. Thesis, University of Wisconsin-Milwaukee.

\reference{StergioulasFriedman} Stergioulas, N. 
   \&  Friedman, J. L. 1995, ApJ {\bf 444}, 306 

\reference{WGW91} Weber, F., Glendenning, N. K. \& Wiegel, M. K. 1991, 
   ApJ, {\bf 373}, 579.  

\reference{YoshidaEriguchi95} Yoshida, S. 
   \& Eriguchi, Y. 1995, ApJ, {\bf 438}, 830

\reference{YoshidaEriguchi97} Yoshida, S. 
   \& Eriguchi, Y. 1997, submitted to ApJ Letters.


\end{references}
\end{document}